\documentclass[12pt]{iopart}
\usepackage{graphicx}
\usepackage{dcolumn}
\usepackage{bm}
\usepackage{caption}
\usepackage{hyperref}
\usepackage{xcolor} 
\usepackage{float}
\usepackage{siunitx}
\usepackage{iopams}
\usepackage[normalem]{ulem}

\def\onedot{$\mathsurround0pt\ldotp$}
\def\cdddot#1{% three dots 
	\mathbin{\vcenter{\baselineskip.67ex
			\hbox{\onedot}\hbox{\onedot}\hbox{\onedot}%
	}}%
}

\newcommand\mab[1]{\textcolor[rgb]{0,0,0}{#1}}

\newcommand\eqref[1]{(\ref{#1})}

\begin{document}
	
	\title[Variational formulation of active nematic fluids: theory and simulation]{Variational formulation of active nematic \mab{fluids}: theory and simulation}
	
	\author{W Mirza$^{1,2}$, A Torres-S\'anchez$^{1,3}$\footnote{Present address:
			Tissue Biology and Disease Modelling Unit, European Molecular Biology Laboratory, Doctor Aiguader 88, Barcelona (08003), Spain.}, 
		G Vilanova$^1$ and Marino Arroyo$^{1,3,4}$}
	
	\address{$^1$ LaC\`aN, Universitat Polit\`ecnica de Catalunya BarcelonaTech, Jordi Girona 1-3 08034 Barcelona, Spain}
	\address{$^2$ Barcelona Graduate School of Mathematics (BGSMath), Campus de Bellaterra, Edifici C
		08193 Bellaterra Barcelona, Spain}
	\address{$^3$ Institute for Bioengineering of Catalonia (IBEC),
		The Barcelona Institute of Science and Technology (BIST), Baldiri Reixac 10-12, 08028 Barcelona Spain}
	\address{$^4$ Centre Internacional de M\`etodes Num\`erics en Enginyeria (CIMNE), 08034 Barcelona, Spain}
	\ead{marino.arroyo@upc.edu}
	\ead{alejandro.torressanchez@embl.es}
	
	\begin{abstract}
%		The structure and dynamics of important biological quasi-two-dimensional systems, ranging from cytoskeletal gels to tissues, are controlled by nematic order, defects and activity. Continuum hydrodynamic descriptions combined with numerical simulations have been used to understand such complex systems, but the physical interpretation of different active nematic models and their applicability to specific systems is often unclear. For instance, most works rely on theories for incompressible liquid crystals but important active 2D nematic systems are compressible due to density variations or turnover. Here, we propose a theoretical and computational framework for possibly compressible and density-dependent 2D active nematic systems. This framework is based on Onsager's variational formalism to irreversible thermodynamics, according to which the dynamics result from a competition between free-energy release, dissipation and activity. We particularize this framework to recover a standard incompressible active nematic model and further formulate an alternative model for density-dependent active nemato-hydrodynamics. We show that the variational principle enables a direct and transparent derivation not only of the governing equations, but also of the finite element numerical scheme. We exercise this model in two representative examples of active nematodynamics relevant to the actin cytoskeleton during wound healing and to the dynamics of confined colonies of elongated cells. 
%\\
\mab{The structure and dynamics of important biological quasi-two-dimensional systems, ranging from cytoskeletal gels to tissues, are controlled by nematic order, flow, defects and activity. Continuum hydrodynamic descriptions combined with numerical simulations have been used to understand such complex systems. The development of thermodynamically consistent theories and numerical methods to model active nemato-hydrodynamics is eased by mathematical formalisms enabling systematic derivations and structured-preserving algorithms. Alternative to classical nonequilibrium thermodynamics and bracket formalisms, here we develop a theoretical and computational framework for active nematics based on Onsager's variational formalism to irreversible thermodynamics, according to which the dynamics result from the minimization of a Rayleighian functional capturing the competition between free-energy release, dissipation and activity. We show that two standard incompressible models of active nemato-hydrodynamics can be framed in the variational formalism, and develop a new compressible model for density-dependent active nemato-hydrodynamics relevant to model actomyosin gels. We show that the variational principle enables a direct and transparent derivation not only of the governing equations, but also of the finite element numerical scheme. We exercise this model in two representative examples of active nemato-hydrodynamics relevant to the actin cytoskeleton during wound healing and to the dynamics of confined colonies of elongated cells. }

	\end{abstract}
	
	%\submitto{\NJP}
	\maketitle
	
	\section{Introduction} 
	
	\mab{Important sub- and supra-cellular biological systems and bioinspired materials can be described as active nematic fluids \cite{marchetti2013, doostmohammadi2018,Bowick2022,BALASUBRAMANIAM2022101897}. Examples include microtubule-kinesin gels \cite{Sanchez:2012aa}, actomyosin gels in vitro \cite{doi:10.1126/science.aao5434}  and in living cells and organisms in the form of dense bundles \cite{lehtimaki2021,maroudas2021,BALASUBRAMANIAM2022101897}, dense bacterial suspensions \cite{wioland2013} or dense colonies of elongated animal cells \cite{duclos2014,duclos2017, guillamat2020}. Active nematic systems are characterized by local nematic order combined with active power input, which couples to nematic order in that contractile or extensile active stresses orient along the nematic direction. Both nematicity and activity can independently induce the self-organization of heterogeneous structures. For instance, passive nematic systems can develop defects, either half-integer such as comets ($+1/2$) and trefoils ($-1/2$) or full-integer ($+1$) such as spirals or asters. Without nematic order, activity can also induce self-organized patterns, such as those resulting from self-reinforcing flows driving density accumulation in contractile cytoskeletal gels \cite{callan2013,Ruprecht:2015aa,hannezo2015}. Acting together, nematicity and activity drive a plethora of dynamical behaviors, \mab{where the nematic phase becomes unstable to bend or splay deformations \cite{simha2002,ramaswamy2007}, leading to defects, which in turn generate active flows and become motile,}  leading to ``low Reynolds number active turbulence'' at high-enough activity \cite{giomi2015,hemingway2016,opathalage2019}. Moreover, the interplay between nematic organization, hydrodynamic flows and density accumulation plays a key role in the self-organization of diverse architectures in actomyosin gels such as dense nematic bundles \cite{salbreux2009,lehtimaki2021,Mirza_2024}.}
	
\mab{Alternative to discrete models of active nematic fluids \cite{bechinger2016}, continuum models in combination with numerical discretization can access the pertinent  mesoscales using fewer degrees of freedom and are amenable to mathematical analysis. Building on the literature of liquid crystals and structured fluids \cite{de1993,Beris1994,Stewart2004}, a number of active nematic continuum models on possibly curved surfaces have been developed and analyzed \cite{simha2002,hatwalne2004,marenduzzo2007,salbreux2009,giomi2014,giomi2015,julicher2018,zhang2020,pearce2020,Axel_Voigt,metselaar2019,Santhosh2020,D2SM00988A}, and numerically approximated with  the Lattice Boltzmann algorithm \cite{marenduzzo2007,cates2009}, the Hybrid Boltzmann algorithm \cite{desplat2001},  and finite element methods \cite{norton2018,Axel_Voigt}. Continuum models of nematic systems can be derived by coarse-graining the microscopic dynamics  \cite{simha2002}. Complementarily, continuum models are often formulated within the framework of irreversible thermodynamics \cite{de2013non}, where phenomenological constitutive relations are developed to be consistent with an entropy production inequality, symmetry and reciprocity relations \cite{salbreux2009,Prost:2015aa,julicher2018, salbreux2022}. Irreversible thermodynamics establishes a clear distinction between dissipative and reversible processes, and therefore it is a natural framework to model active processes, which are neither dissipative nor reversible in general.
In an attempt to further structure the theory of irreversible thermodynamics for flowing systems, the Poisson bracket formalism of Hamiltonian reversible mechanics has been extended to nonequilibrium systems by introducing a dissipative bracket, leading to the GENERIC \cite{Grmela1984,Morrison1984,Kaufman1984} and the single-generator bracket \cite{Beris1990,Beris1994} formalisms. The bracket structure of the dynamics combines symplectic and gradient flow components, and guarantees by construction thermodynamic consistency, that is energy conservation and entropy production. Hydrodynamic theories of active nematic models in the literature are often based on the Beris-Edwards bracket formalism \cite{marenduzzo2007,giomi2014,giomi2015,zhang2020,pearce2020,Axel_Voigt,metselaar2019,Santhosh2020,D2SM00988A}. Besides explicitly revealing the mathematical structure of the dynamical equations, bracket formalisms facilitate the systematic formulation of thermodynamically-consistent models of complex systems, and provide a framework for structure-preserving numerical algorithms \cite{Suzuki2017}, which at a discrete level preserve exactly qualitative properties of the continuum system, and for physics-informed machine learning of nonequilibrium systems \cite{HERNANDEZ2021109950,Zhang2022}.}

\mab{Soft matter and biological systems often operate at low Reynolds and isothermal conditions. In such situations, it is possible to describe irreversible thermodynamics using a variational formalism, the so-called Onsager's variational principle, in which the dynamics minimize instantaneously a Rayleighian functional \cite{doi2011,doi2012,Mielke2012,peletier2014variational,arroyo2018}. This variational framework is rooted in early work by Rayleigh and Onsager \cite{rayleigh1873,Onsager1931,PhysRev.38.2265}, and despite the contentious standing of early attempts to unveil a nonequilibrium optimality principle \cite{Beris2024}, it constitutes a \emph{bona fide} formalism to irreversible thermodynamics that can be justified as emerging from microscopic stochastic processes \cite{peletier2014variational,D0SM02076A,Mielke2014,mielke2016generalization,Kraaij2020,Montefusco2021}  and that is obtained as a suitable limit of bracket formalisms \cite{Esen2022,Beris2024}. It recapitulates a wide variety of soft matter systems models involving hydrodynamics, diffusion, reactions and elasticity reviewed in \cite{D0SM02076A}. It also provides a structured modeling framework to systematically formulate thermodynamically consistent theories of nonlinear multi-physics and possibly active systems in bulk or interfacial moving domains such as those governing the reshaping of biomembranes \cite{rahimi2012,Callan-Jones2016,Tozzi_2019}, active gels of the actin cytoskeleton \cite{PhysRevE.94.012403,torres2019}, or the patterning of growing tissues \cite{Ackermann2023,CICCONOFRI2024105773}.}

\mab{In Onsager's formalism, the dynamics minimize a Rayleighian functional where free-energy release, dissipation and (possibly active) power input compete, see \ref{Ons_Form}.
%\ats{[This idea of competition is a bit strange to me --it's in several places in the article--; this is more clear to me: In Onsager's formalism, the dynamics minimize a Rayleighian functional that measures the difference between the true energy dissipation, which is obtained by adding the rate of change of the free energy and the (possibly active) power input, and a dissipation potential, which encodes the constitutive laws for the dissipative mechanisms acting in the system, see \ref{Ons_Form}.]}. 
The first-order optimality conditions yield the evolution equations, including those corresponding to generalized force-balance, as well as energy balance. The second-order optimality conditions generalize Onsager's reciprocal relations and guarantee the entropy production inequality under mild assumptions on the dissipation potential. The first-order conditions can be viewed as a generalized gradient flow with possibly non-quadratic dissipation \cite{EDELEN1972481,PhysRevE.47.351,mielke2016generalization}, which neglects the inertial Hamiltonian flow component of bracket formalisms. Thus, Onsager's formalism trades-off generality for mathematical structure (a minimum principle). When dealing with systems involving tensor-valued fields or non-Euclidean domains, the objectivity of the theory is guaranteed in Onsager's formalism by the objectivity of a single scalar functional, the Rayleighian, which greatly simplifies modeling. The variational structure further enables the development of reduced theories based on ansatz \cite{Doi2015,zhang2020},  the development of structure-preserving discretizations that harness optimization algorithms \cite{torres2019,Hu2024}, or the integration with machine learning methods \cite{Huang2022}. However, to our knowledge, only one specific model of active nemato-hydrodynamics has been formulated within Onsager's formalism to obtain approximate solutions \cite{zhang2020}.}

\mab{The objective of the present work is two-fold. First, we aim at  establishing a general theoretical framework for active nemato-hydrodynamics based on Onsager's variational formalism, granting a transparent and systematic procedure to develop models and numerical discretizations for such complex systems. Second, we aim at using the general framework to develop a theoretical description of the actomyosin cytoskeleton, focusing on the emergence of dense nematic phases. At time-scales of minutes and up, the elastic stresses in the actin cytoskeleton can be assumed to have dissipated and the system can be understood as a compressible, active and viscous fluid gel with orientational order \cite{salbreux2012,BALASUBRAMANIAM2022101897}. The dynamics of the actomyosin cytoskeleton is often characterized by strong density variations and compressible flows at low Reynolds numbers, for instance during the prominent advective instability involving self-reinforcing flows in contractile gels \cite{callan2013,Ruprecht:2015aa,hannezo2015}, which can elicit the self-assembly of dense nematic structures \cite{salbreux2009}. These systems should therefore be described as compressible and density-dependent active nematic fluids. However, and despite a vast literature, such a model has not been developed to the best of our knowledge. Indeed, models for dry aligning dilute active matter \cite{annurev_Chate,C6SM00268D} can describe large density fluctuations and compressible flows, but fail to describe the wet physics of cytoskeletal hydrodynamics. Likewise, neither common nemato-hydrodynamic models ignoring density fields, incompressible \cite{marenduzzo2007,giomi2015,julicher2018,zhang2020,pearce2020,Axel_Voigt,metselaar2019,Santhosh2020} or compressible \cite{C6SM01493C,PhysRevE.106.054610}, nor models accounting for the density of active particles suspended in an incompressible fluid \cite{simha2002,hatwalne2004,giomi2014,D2SM00988A}, can describe the self-reinforcing flows characteristic of actomyosin gels.}

\mab{The paper is organized as follows. In Sections \ref{sec_1} and \ref{sec:Onsager}, we develop a general theory for continuum active density-dependent nemato-hydrodynamics based on Onsager's variational formalism (Sections \ref{sec_1} and \ref{sec:Onsager}). To highlight that Onsager's formalism is compatible with other formalisms of irreversible thermodynamics, in Section \ref{sec_2_bis} we show that two standard incompressible models for active nemato-hydrodynamics admit a variational derivation. In Sec.~\ref{sec_3}, we develop a minimal compressible and density-dependent model for active nemato-hydrodynamics pertinent to actomyosin gels. In \ref{sec_4}, we develop a finite element formulation to approximate the theoretical model using Onsager's variational formalism. In Sec.~\ref{sec_5}, we use our model and algorithm to explore the active self-organization of nematic architectures in two biologically  relevant situations, namely the assembly of an actomysin contractile ring leading to wound healing in large egg cells \cite{benink2000,mandato2001} and defect dynamics in confined populations of elongated cells \cite{duclos2014,duclos2017}. These computational studies highlight the role of compressibility in the self-organization of active nematic fluids. Finally, in Sec.~\ref{summary}, we provide a summary and outlook of our work.}

	\section{Description of a 2D active nematic gel} \label{sec_1}
	
	We consider a planar thin sheet of a nematic fluid. We describe this system with its areal density field $\rho(\bm{x},t)$ and with a \mab{symmetric and traceless tensor field $\bm{q}(\bm{x},t)$ characterizing local order called nematic tensor field. For interpretation purposes, we note that in 2D the nematic tensor can be represented as 
	\begin{equation} 
		\label{eq:nematic_tensor}
		\bm{q}(\bm{x},t) =S(\bm{x},t) \left[\bm{n}(\bm{x},t) \otimes \bm{n}(\bm{x},t) - \frac{1}{2}\bm{I}\right] ,
	\end{equation}
	where $\bm{I}$ is the 2D identity tensor, $\bm{n}$ is a unit vector representing the average nematic alignment, $\otimes$ is the dyadic or outer product, and $S=\sqrt{2 q_{ab}q_{ab}}>0$} is the nematic order parameter, which measures the strength of the nematic alignment about $\bm{n}$. $q_{ab}$ denote the components of $\bm{q}$ in a Cartesian basis and we follow Einstein's summation convention for repeated indices. The density and nematic fields $\rho(\bm{x},t)$ and $\bm{q}(\bm{x},t)$ are the state variables, denoted by $X(t)$ in \ref{Ons_Form}.
	
We denote by $\bm{v}$ the hydrodynamic velocity field and decompose its gradient into a symmetric and an antisymmetric parts
	\begin{equation} \label{gradv}
		\nabla \bm{v}  = \bm{d} + \bm{w},
	\end{equation}
	where 
	\begin{eqnarray}
		\label{eq:rate-of-deformation}
		{d}_{ab} &= \frac{1}{2}\left(\nabla_b {v}_a +\nabla_a{v}_b\right),\\ 
		\label{eq:spin}
		{w}_{ab} &= \frac{1}{2}\left(\nabla_b {v}_a -\nabla_a{v}_b\right). 
	\end{eqnarray}
	The tensor $\bm{d}$ characterizes the rate of deformation of a differential of volume of the material and it is usually referred to as the rate-of-deformation tensor. This tensor plays an important role in different aspects of the theory. For instance, $\text{tr}\bm{d}=\nabla \cdot \bm{v}$ measures local area changes and hence contributes to the rate of accumulation or dilution of density, and viscous dissipation in the fluid is formulated in terms of $\bm{d}$ as discussed later. While $\text{tr}\bm{d}$ describes the isotropic part of the rate of deformation, the deviatoric part is described by the traceless tensor
	\begin{equation}
		\bm{d}^{\rm dev} = \bm{d} - \frac{{\rm tr}\bm{d}}{2} \bm{I}.
	\end{equation}
	%and its deviatoric part of $\bm{d}^{\rm dev} = \bm{d}  - \frac{1}{2}(\text{tr}\bm{d})\bm{I}$ identifies the shear rate leading to energy dissipation.
	%\ats{and $\eta (|\bm{d}|^2+ (\tr \bm{d})^2)$ is the energy dissipation for a 3D incompressible Newtonian thin sheet, where $\eta$ is the shear viscosity.} 
	The tensor $\bm{w}$ describes the local rate of rotation induced by the flow, possibly leading to rotation of the nematic alignment, and it is referred to as the spin tensor; note that in 2D it can be represented with a single scalar $\bm{w}=w\bm{\epsilon}$, where $\bm{\epsilon}$ is the Levi-Civita tensor. We denote the gradient of the spin as 
	\begin{equation} 
		\label{zeta}
		\bm{\zeta} = \nabla w \, .
	\end{equation} 
	We characterize the rate of change of $\bm{q}$ with the Jaumann derivative \cite{de1993}
	\begin{equation}
		\label{eq:Jaumann}
		\widehat{\bm{q}}= \dot{\bm{q}}  + \bm{q}\,\bm{w} -  \bm{w}\,\bm{q},
	\end{equation}
	or in components as
	\begin{equation} \label{jaumann_detivative_def}
		\widehat{q}_{ab} = \dot{q}_{ab}  + q_{ac} w_{cb} - w_{ac} q_{cb}.
	\end{equation}
	where $ \dot{q}_{ab} = \partial_t q_{ab} + v_c \nabla_c q_{ab}$ is the total time derivative of the nematic order tensor. $\widehat{\bm{q}}$ measures the rate of change of $\bm{q}$ viewed by an observer that flows and rotates with $\bm{v}$, i.e.~it can be geometrically viewed in terms of Lie derivatives \cite{marsden1994}; thus, $\widehat{\bm{q}}$ is zero if $\bm{q}$ is advected and rotated by the flow without any further rearrangements of the nematic field. The fields $\bm{v}$ and $\widehat{\bm{q}}$ define the process variables $V$ of our system.
	
	The process operator relating process variables and time-derivatives of the state variables, denoted by $\partial_t X = P(X)V$ in \ref{Ons_Form}, is specified by Eq.~(\ref{eq:Jaumann}) along with the mass conservation equation for $\rho$ given by 
	\begin{equation}
		\label{eq:balance_mass}
		\dot{\rho} + \rho {\rm tr}~\bm{d} = r ,
	\end{equation}
	where $\dot{\rho}(\bm{x},t) = \partial_t \rho(\bm{x},t) + \bm{v}(\bm{x},t)\cdot\nabla\rho(\bm{x},t)$ is the material time-derivative of $\rho$. The second term characterizes the dilution (or compaction) of $\rho$ caused by the rate of change of local area, and $r$ is the rate of change of density not explained by the flow $\bm{v}$, which typically results from chemical reactions and diffusion. Although the diffusive fluxes and reaction rates leading to $r$ can be deduced from an extended Onsager variational formalism \cite{Mielke2012}, here we focus on active nemato-hydrodynamics and hence assume $r$ as given.

	\begin{figure}
		\centering
		\includegraphics[width=0.9\textwidth]{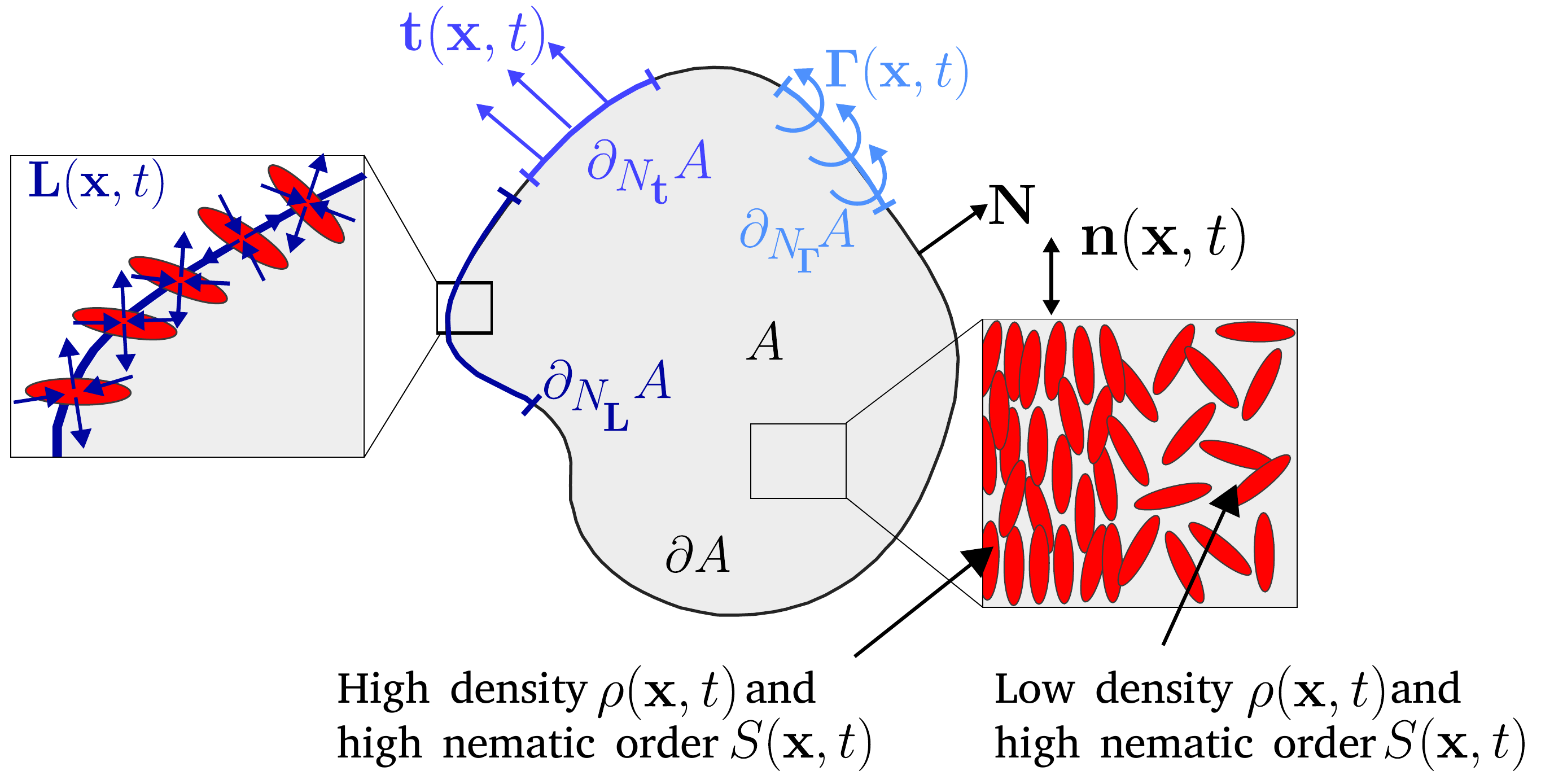}
		\caption{\label{fig0}  \textbf{Schematic of a compressible active nematic gel in 2D}. The system occupies a region $A$ with boundary $\partial A$ and unit outer boundary normal $\bm{N}$. Traction $\bm{t}$, moment $\bm{\Gamma}$ and microscopic moment $\bm{L}$ act on the  boundaries $\partial_{N_{\bm{t}}} A $, $\partial_{N_{\bm{\Gamma}}} A $ and $\partial_{N_{\bm{L}}} A $, respectively. $\bm{L}$ is symmetric and traceless tensor represented with two orthogonal pairs of diverging/converging arrows of equal length and $\bm{\Gamma} = \Gamma \bm{\epsilon}$, with $\bm{\epsilon}$ the Levi-Civita tensor, is an antisymmetric tensor represented as a torque of magnitude $\Gamma$.	The active nematic system exhibits spatiotemporal variations in density $\rho(\bm{x},t)$ and the nematic order tensor $\bm{q}(\bm{x},t)$, parametrized by the nematic order parameter $S(\bm{x},t)$ and the average molecular orientation $\bm{n}(\bm{x},t)$, Eq.~(\ref{eq:nematic_tensor}), represented with a double-headed arrow.  	}
	\end{figure}
	
	\section{Variational formalism for a generic active nematic gel} \label{sec:Onsager}
	
	\subsection{Free-energy, dissipation and power input and Rayleighian functionals}
	
	We derive next the governing equations of an active nematic gel at low Reynolds numbers, occupying a region $A \subset \mathbb{R}^2$ with boundary $\partial A$. The unit outward normal to $\partial A$ is denoted by $\bm{N}$.  
	We assume Dirichlet boundary conditions for $\bm{v}$, $\bm{w}$ and $\widehat{\bm{q}}$ in  subsets of the boundary denoted by  $\partial_{D_{\bm{v}}} A $, $\partial_{D_{\bm{w}}} A $ and $\partial_{D_{\widehat{\bm{q}}}} A $, respectively.
	At Neumann boundaries, denoted by $\partial_{N_{\bm{t}}} A $, $\partial_{N_{\bm{\Gamma}}} A $ and $\partial_{N_{\bm{L}}} A $, we prescribe the traction vector, $\bm{t}$, the antisymmetric torque tensor, $\bm{\Gamma}$, and the generalized force power-conjugate to $\widehat{\bm{q}}$ represented by a symmetric traceless tensor, $\bm{L}$. Dirichlet and Neumann boundaries are pairwise complementary, e.g.~$\partial_{D_{\bm{v}}} A \cup \partial_{N_{\bm{t}}} A = \partial A$ and $\partial_{D_{\bm{v}}} A \cap \partial_{N_{\bm{t}}} A = \emptyset$. See Fig.~\ref{fig0} for an illustration.

	To derive the governing equations, we follow the systematic procedure given by Onsager's variational formalism summarized in \ref{Ons_Form}. We postulate generic forms for the free-energy, dissipation and power input functionals as
	\begin{eqnarray}
		\mathcal{F}\left[\rho,\bm{q}\right] &=& \int_A f(\bm{q},\nabla\bm{q}) \rho dA, \label{eq:free_energy}\\
		\mathcal{D}\left[\bm{v},\widehat{\bm{q}}; \rho,\bm{q}\right] &=& \int_A d(\bm{v},\bm{d},\bm{w},\bm{\zeta},\widehat{\bm{q}};\rho,\bm{q}) \rho \,dA,  \label{eq:dissipation}\\
		\mathcal{P}\left[\bm{v},\widehat{\bm{q}}; \rho,\bm{q}\right] &=& \int_A p(\bm{v},\bm{d},\bm{w},\bm{\zeta},\widehat{\bm{q}};\rho,\bm{q}) \rho \,dA \nonumber\\
		& &~~- \int_{\partial_{N_{\bm{t}}} A} \bm{t} \cdot \bm{v} \,dl -  \int_{\partial_{N_{\bm{\Gamma}}} A} \bm{\Gamma}:\bm{w} \,dl -  \int_{\partial_{N_{\bm{L}}} A} \bm{L}:\widehat{\bm{q}} \,dl, \label{eq:power}
	\end{eqnarray}
	where $f$, $d$ and $p$ are the free-energy, dissipation, and power input densities per unit mass. The free energy depends on the state variables $\rho$ and $\bm{q}$. The dissipation and power inputs depend on process variables $\bm{v}$, $\widehat{\bm{q}}$, and might also depend parametrically on the state variables. Examining Eq.~(\ref{eq:Jaumann}), it is clear that we could have alternatively chosen $V = \left(\bm{v}, \dot{\bm{q}}\right)$ or $V = \left(\bm{v}, \partial_t{\bm{q}}\right)$ as process variables, leading to different forms of the Euler-Lagrange equations. The central functional in Onsager's variational formalism is the Rayleighian $\mathcal{R} = {d{\mathcal{F}}}/{dt} + \mathcal{D} + \mathcal{P}$. We compute next an expression of the rate of change of the free energy.

	%%%%%%%%%%%%%%%%%%%%%%%%%%%%%%%%%%%%%%%%%%%%%%%%%%%%%%%%%%%%%%%%%%%%%%%%
	
	\subsection{Rate of change of the free energy functional}\label{app:rate_of_change_energy}
	
Applying Reynolds transport theorem to the integral in Eq.~(\ref{eq:free_energy}), we obtain 
	\begin{eqnarray} 
		\frac{d{\mathcal{F}}}{dt} &= \int_A \left[\partial_tf  \rho + f \partial_t{\rho} \right] dA + \int_{\partial A} f\rho \bm{v} \cdot\bm{N} ~dl = \int_A \left[\partial_tf  \rho + f \partial_t{\rho}  + \nabla\cdot\left(f\rho \bm{v}\right) \right] dA  \nonumber\\
		&=\int_A \left[\dot{f}  \rho + f \dot{\rho}  + f\rho {\rm tr}\bm{d} \right] dA. \label{eq:change_of_free_energy}
	\end{eqnarray}
\mab{The rate of change of the free energy functional competes in the Rayleighian functional with the dissipation and power functionals $\mathcal{D}$ and $\mathcal{P}$. Therefore, we need to express the integrand of the expression above as a function of $(\bm{v},\bm{d},\bm{w},\bm{\zeta},\widehat{\bm{q}};\rho,\bm{q})$, see Eqs.~(\ref{eq:dissipation},\ref{eq:power}). Lengthy but otherwise direct mathematical manipulations allow us to express the material time-derivative of $f$ as 	
	\begin{eqnarray}
		\dot{f} = \frac{\partial f}{\partial q_{ab}}\widehat{q}_{ab} + \frac{\partial f}{\partial \nabla_c q_{ab}}\left( \nabla_c \widehat{q}_{ab} - d_{dc} \nabla_d q_{ab} + 2\epsilon_{ad} q_{db}  \zeta_c\right).
		\label{eq:final_rate_of_free_energy}
	\end{eqnarray}
See \ref{App:dFdt} for a detailed derivation. To obtain this expression, a crucial step is to enforce that $f$ should obey the principle of material frame indifference \cite{Stewart2004}, according to which it should remain invariant under rigid body motions.}	Plugging this expression in Eq.~(\ref{eq:change_of_free_energy}), we obtain
	\begin{eqnarray} 
		\frac{d{\mathcal{F}}}{dt} =  \int_{A}  \left[  f\frac{\dot{\rho}}{\rho}  + f  {\rm tr}\bm{d} +\frac{\partial  f}{\partial \bm{q}} : \widehat{\bm{q}}  + \frac{\partial f}{\partial \nabla_c q_{ab}}\left( \nabla_c \widehat{q}_{ab} - d_{dc} \nabla_d q_{ab} + 2\epsilon_{ad} q_{db}  \zeta_c\right) \right] \rho dA.\label{eq:dotF}
	\end{eqnarray}

	\subsection{System  Rayleighian}
	
	For clarity of our derivation, we consider the fields $\dot{\rho}$, $\bm{d}$, $\bm{w}$ and $\bm{\zeta}$ as independent variables, and enforce the kinematic and conservation relations relating them through Lagrange multipliers. Thus, the Rayleighian has the form
	\begin{eqnarray} 
		\mathcal{R}\left[\dot{\rho},\bm{v},\bm{d},\bm{w},\bm{\zeta},\widehat{\bm{q}};\rho,\bm{q}\right] = & \frac{d{\mathcal{F}}}{dt}[\dot{\rho},\bm{d},\bm{\zeta},\widehat{\bm{q}}; \rho,\bm{q}] + \nonumber
		\mathcal{D}\left[\bm{v},\bm{d},\bm{w},\bm{\zeta},\widehat{\bm{q}}; \bm{q},\rho\right]   \\ & +\mathcal{P}\left[\bm{v},\bm{d},\bm{w},\bm{\zeta},\widehat{\bm{q}}; \bm{q},\rho\right], 	\label{eq:Rayleighian}
	\end{eqnarray}
	and the governing equations can then be obtained according to Onsager's variational principle by minimizing it with respect to the extended process variables $\left(\dot{\rho}, \bm{v},\bm{d},\bm{w},\bm{\zeta},\widehat{\bm{q}}\right)$ subject to  kinematic and mass conservation constraints expressed by the functional
	\begin{eqnarray} 
		\mathcal{Q}[\varrho, & \bm{\sigma}^{\rm s}, \bm{\sigma}^{\rm a},  \bm{m}, \dot{\rho},\bm{v},\bm{d},\bm{w},\bm{\zeta},\widehat{\bm{q}};\rho,\bm{q}] =
		\int_A \left\{ \varrho  \left[\dot{\rho} + \rho{\rm tr}~\bm{d}- r\right] \vphantom{\frac{1}{2}}\right.\nonumber\\
		& +\bm{\sigma}^{\rm s} :\left[\bm{d} - \frac{1}{2}\left(\nabla \bm{v} + \left(\nabla\bm{v}\right)^T\right)\right]  %\nonumber\\& 
		+\bm{\sigma}^{\rm a} :\left[\bm{w} - \frac{1}{2}\left(\nabla \bm{v} - \left(\nabla\bm{v}\right)^T\right)\right]  \nonumber\\
		& +\left. \vphantom{\frac{1}{2}}\bm{m} \cdot \left[\bm{\zeta} - \nabla w\right] \right\}dA, \label{eq:constraints}
	\end{eqnarray}
	where $\varrho $ is the Lagrange multiplier imposing balance of mass; $\bm{\sigma}^{\rm s}$, a symmetric tensor, and $\bm{\sigma}^{\rm a}$, an antisymmetric tensor, are the Lagrange multipliers imposing the definitions of the rate-of-deformation and spin tensors; and $\bm{m}$, a vector, is the Lagrange multiplier imposing the definition of the gradient of the spin.  We thus form the  Lagrangian as 
	\begin{eqnarray} 
		\mathcal{L}[\varrho, \bm{\sigma}^{\rm s}, \bm{\sigma}^{\rm a},\bm{m},\dot{\rho},\bm{v},\bm{d},\bm{w},\bm{\zeta},\widehat{\bm{q}};&\rho,\bm{q}] = \mathcal{R}[\dot{\rho},\bm{v},\bm{d},\bm{w},\bm{\zeta},\widehat{\bm{q}};\rho,\bm{q}] \nonumber \\
		&-\mathcal{Q}[\varrho, \bm{\sigma}^{\rm s}, \bm{\sigma}^{\rm a},\bm{m}, \dot{\rho},\bm{v},\bm{d},\bm{w},\bm{\zeta},\widehat{\bm{q}};\rho,\bm{q}].	\label{eq:Lagrangian} 
	\end{eqnarray}
	
	\subsection{Balance equations and constitutive equations as optimality conditions}\label{sec_2}
	
	We examine next the first order optimality conditions. Making $\mathcal{L}$ stationary with respect to $\widehat{\bm{q}}$ and integrating by parts leads to
	\begin{eqnarray} 
		0&=&\int_A \left[ \left(\frac{\partial  f}{\partial \bm{q}} + \frac{\partial d}{\partial \widehat{\bm{q}}}+ \frac{\partial p}{\partial \widehat{\bm{q}}} \right) : \delta \widehat{\bm{q}} + \frac{\partial f}{\partial \nabla_c q_{ab}} \nabla_c \delta \widehat{q}_{ab} \right] \rho dA - \int_{\partial_{N_{\bm{L}}}A} \bm{L}:\delta\widehat{\bm{q}} dl \nonumber\\
		&=&\int_A \left[\rho \left(\frac{\partial  f}{\partial {q}_{ab}} + \frac{\partial d}{\partial \widehat{{q}}_{ab}}+ \frac{\partial p}{\partial \widehat{{q}}_{ab}} \right) - \nabla_c \left(\rho \frac{\partial f}{\partial \nabla_c {q}_{ab}} \right)   \right] \delta \widehat{{q}}_{ab}  dA \nonumber\\
		&&+ \int_{\partial_{N_{\bm{L}}}A} \left(\rho \frac{\partial f}{\partial \nabla_c {q}_{ab}} N_c -{L}_{ab}\right)\delta\widehat{{q}}_{ab} dl, \label{eq:var_q}
	\end{eqnarray}
	for all admissible traceless and symmetric variations $\delta\widehat{{q}}_{ab}$ that vanish on $\partial_{D_{\widehat{\bm{q}}}} A $. Localizing this equation, we obtain
	\begin{eqnarray} 
		\rho \left(\frac{\partial d}{\partial \widehat{{q}}_{ab}}+ \frac{\partial p}{\partial \widehat{{q}}_{ab}} \right) - h_{ab} = 0  \label{eq:balance_q} \qquad & \text{in } A ,\\
		\rho \frac{\partial f}{\partial \nabla_c {q}_{ab}} N_c = {L}_{ab} \qquad & \text{on } \partial_{N_{\bm{L}}} A, \label{eq:balance_q_bd} 
	\end{eqnarray}
where we have introduced the functional derivative of the free-energy with respect to the nematic tensor
\begin{equation}
\label{nem_field}
h_{ab} = -\frac{\delta \mathcal{F}}{\delta {q}_{ab}} = -\rho \frac{\partial f}{\partial {q}_{ab}} + \nabla_c \left(\rho \frac{\partial f}{\partial \nabla_c {q}_{ab}}\right), \label{molecular_field}
\end{equation}
\mab{referred-to in the literature as the molecular field. Any hydrostatic component in Eq.~(\ref{eq:balance_q}) does not perform power against the traceless variations $\delta\widehat{{q}}_{ab}$ in Eq.~(\ref{eq:var_q}), and therefore, without loss of generality, we can constrain $d$, $p$ and $f$ so that each of the terms in Eq.~(\ref{eq:balance_q}) are traceless.} Equations~(\ref{eq:balance_q},\ref{eq:balance_q_bd}) express balance of generalized forces power-conjugate to $\widehat{\bm{q}}$ in the bulk and at the boundary. 
	
	Variations with respect to $\dot{\rho}$ lead to  $\varrho  = f$.
	Using this result, stationarity of $\mathcal{L}$ with respect to $\bm{d}$ provides a definition for $\bm{\sigma}^{\rm s}$:
	\begin{eqnarray} 
		\sigma^{\rm s}_{ab} = \rho\left[-\frac{1}{2} \left(\frac{\partial  f}{\partial \nabla_b q_{dc}} \nabla_a q_{dc}+ \frac{\partial  f}{\partial \nabla_a q_{dc}} \nabla_b q_{dc}\right) +  \frac{\partial  d}{\partial d_{ab}}+  \frac{\partial  p}{\partial d_{ab}}\right].
		\label{eq:sym_stress}
	\end{eqnarray}
	%where we have used the fact that variations with respect to $\dot{\rho}$ lead to $\varrho  = -f$. 
	Variations with respect to $\bm{\zeta}$ lead to
	\begin{equation}
		\label{eq:moment}
		m_c =\rho  \left(2 \epsilon_{ad} \frac{\partial  f}{\partial \nabla_c q_{ab}} q_{db} + \frac{\partial  d}{\partial \zeta_c} + \frac{\partial  p}{\partial \zeta_c}\right).
	\end{equation}
	Introducing 
	\begin{equation}
		\bm{\omega} = - \rho \left(\frac{\partial  d}{\partial \bm{w}} + \frac{\partial  p}{\partial \bm{w}}\right),
	\end{equation}
	stationarity of $\mathcal{L}$ with respect to $\bm{w}$ leads to
	\begin{eqnarray} 
		0&=\int_A \left[-\bm{\sigma}^{\rm a} :\delta\bm{w} + \frac{1}{2} {m}_c \epsilon_{ab} \nabla_c \delta w_{ab}  - \bm{\omega} :\delta \bm{w} \right] dA - \int_{\partial_{N_{\bm{L}}}A} \bm{\Gamma}:\delta\bm{w} dl \\
		&=\int_A -\left[\bm{\sigma}^{\rm a}  + \frac{1}{2} \left(\nabla \cdot \bm{m}\right) \bm{\epsilon} +  \bm{\omega}  \right] : \delta\bm{w} \;dA + \int_{\partial_{N_{\bm{L}}}A} \left[\frac{1}{2}\left(\bm{m}\cdot\bm{N}\right) \bm{\epsilon}- \bm{\Gamma}\right]:\delta\bm{w}\; dl, \nonumber
	\end{eqnarray}
for arbitrary antisymmetric variations $\delta\bm{w}$ that vanish on $\partial_{D_{\bm{w}}} A$. Localization leads to 
	\begin{eqnarray} 
		\label{sigma_a}
		\bm{\sigma}^{\rm a} + \frac{1}{2}\left(\nabla\cdot\bm{m}\right)\bm{\epsilon} +  \bm{\omega} =  \bm{0}& \qquad \text{in } A, \\ 
		\frac{1}{2}(\bm{m}\cdot\bm{N})\bm{\epsilon} =  \bm{\Gamma} & \qquad \text{on } \partial_{N_{\bm{\Gamma}}} A, 	\label{eq:balance_angular_momentum}
	\end{eqnarray}
	which is a statement of balance of angular momentum, with $\bm{m}$ playing the role of the moment in a Cosserat theory \cite{cosserat1896theorie} and $\bm{\omega}$ of body torques. Equation \eqref{sigma_a} provides a definition for $\bm{\sigma}^{\rm a}$. Combining Eqs.~(\ref{eq:balance_angular_momentum}, \ref{eq:balance_q_bd}) and the  definition for $\bm{m}$ in Eq.~\eqref{eq:moment}, we find the following boundary condition
    \begin{equation}
        \frac{1}{2} \rho \frac{\partial  (d+p)}{\partial \zeta_c} N_c \epsilon_{ab} = \Gamma_{ab} - \left(L_{ae} q_{be} - L_{be} q_{ae}\right) \qquad \text{on } \partial_{N_{\bm{L}}} A \cap \partial_{N_{\bm{\Gamma}}} A, 	\label{eq:border_xi_combined}.
    \end{equation}
	If $d$ and $p$ are independent of $\bm{\zeta}$, this equation shows that  $\bm{\Gamma}$ and $\bm{L}$ cannot be chosen independently; a  generalized force acting on nematic alignment determines the  mechanical torque at the boundary.  In this case, it is necessary that $\partial_{N_{\bm{\Gamma}}}A=\partial_{N_{\bm{L}}} A$ and $\partial_{D_{\hat{\bm{q}}}}A=\partial_{D_{\bm{w}}} A$.
	
Introducing $\bm{\sigma} = \bm{\sigma}^{\rm s} + \bm{\sigma}^{\rm a}$, stationarity of $\mathcal{L}$ with respect to $\bm{v}$ leads to 
	\begin{eqnarray} 
		0 &= \int_A \left[\bm{\sigma}:\nabla\delta\bm{v} - \delta\bm{v}\cdot\bm{f} \right]dA - \int_{\partial_{N_{\bm{t}}} A}  \bm{t} \cdot \delta\bm{v} dl \nonumber\\
		&=\int_A \left[-\nabla\cdot\bm{\sigma} - \bm{f} \right] \cdot \delta\bm{v}\;dA + \int_{\partial_{N_{\bm{t}}} A} \left(\bm{\sigma}\cdot \bm{N} -\bm{t} \right)\cdot \delta\bm{v} \;dl.	\label{eq:weak_v}
	\end{eqnarray}
	for arbitrary $\delta\bm{v}$ that vanish on $\partial_{D_{\bm{v}}} A$ where we have introduced  
	\begin{eqnarray} 
	\bm{f}=-\rho \frac{\partial  (d+p)}{\partial \bm{v}}. \label{force_dens}
	\end{eqnarray} 
Localizing this equation, we find the statement of balance of linear momentum in the absence of inertia
	\begin{eqnarray} 
		\nabla\cdot\bm{\sigma} + \bm{f} = \bm{0} \qquad & \text{in } A, \label{eq:balance_linear_momentum}\\
		\bm{\sigma}\cdot\bm{N} = \bm{t} \qquad & \text{on } \partial_{N_{\bm{t}}} A.	\label{eq:balance_linear_momentum_bd}
	\end{eqnarray}
	We can hence identify $\bm{\sigma}$ as the Cauchy stress tensor, with $\bm{\sigma}^{\rm s}$ and $\bm{\sigma}^{\rm a}$  its symmetric and antisymmetric parts, and $\bm{f}$ as a force density of dissipative and active/external origin.  Finally, variations with respect to $\varrho $, $\bm{\sigma}^{\rm s}$ and $\bm{\sigma}^{\rm a}$ lead to balance of mass in Eq.~(\ref{eq:balance_mass}) and the definitions of the rate-of-deformation and spin tensors, Eqs.~(\ref{eq:rate-of-deformation}) and~(\ref{eq:spin}).

\mab{We can further simplify the theory by combining Eqs.~(\ref{eq:sym_stress},\ref{eq:moment},\ref{sigma_a}) and invoking frame indifference of $f$, see \ref{App:stress}, to obtain a more explicit expression of the total stress tensor as 
	\begin{eqnarray}
		{\sigma}_{ab} & = &  -\rho\frac{\partial  f}{\partial \nabla_b q_{dc}} \nabla_a q_{dc}  +   q_{ad}  h_{bd} - q_{bd} h_{ad}   \nonumber  \\ 
		&& + \rho \frac{\partial (d+p)}{\partial d_{ab}} + \rho \frac{\partial (d+p)}{\partial w_{ab}} -  \frac{1}{2} \nabla_c \left(\rho\frac{\partial (  d+p)}{\partial \zeta_c} \right) \epsilon_{ab}. \label{total_stress}
	\end{eqnarray}
In summary, the system of partial differential equations governing the dynamics of the unknown fields $\rho$, $\bm{v}$ and $\bm{q}$ are the mass balance equation, Eq.~(\ref{eq:balance_mass}), the balance of linear momentum in Eq.~(\ref{eq:balance_linear_momentum}) with the constitutive relations in Eqs.~(\ref{force_dens},\ref{total_stress}), and the generalized nematic force balance in Eq.~(\ref{eq:balance_q}) with the constitutive relation in Eq.~(\ref{molecular_field}).}

	\section{Recovery of two standard models for active nemato-hydrodynamics} \label{sec_2_bis}
	
\mab{We show next that the low Reynolds limit of two different incompressible models for active nemato-hydrodynamics used in the literature can be cast in the general variational framework developed in the previous Section. The first of these models is derived within the classical framework of irreversible thermodynamics \cite{julicher2018}, whereas the second one is a widely used active version of the Beris-Edwards model for liquid crystals resulting from a single-generator bracket formalism \cite{Beris1994}.}
	
	In both cases, we consider an incompressible gel (${\rm tr}\bm{d} = \nabla_a v_a =0$) with initial uniform density and $r=0$. As a result, density remains uniform and can be ignored by considering free-energy, dissipation and power input densities per unit area. %We further assume that 
	%\begin{equation}
	%f(\bm{q},\nabla\bm{q}) = f_0(\bm{q}) + \frac{L}{2}\nabla_c q_{ab} \nabla_c q_{ab},
	%\end{equation}
	%where $L>0$ is the Frank constant. 
	Because of incompressibility, the rate-of-deformation tensor is traceless and  $\bm{d}^{\rm dev} = \bm{d}$.

	\subsection{Variational derivation of a first model for active nematic fluids} \label{form_IT}
	
	We define a quadratic dissipation functional by its density
	\begin{equation}
		\label{diss_inc}
		d(\bm{v},\bm{d},\widehat{\bm{q}}) = \eta \vert\bm{d}^{\rm dev}\vert^2 +\frac{\eta_{\text{rot}}}{2}  \left|\widehat{\bm{q}}\right|^2+ \beta  \bm{d}^{\rm dev}:\widehat{\bm{q}}  + \frac{\gamma}{2} \left|\bm{v}\right|^2,
	\end{equation}
	where $\eta>0$ is the shear viscosity, $\eta_{\text{rot}}>0$ a viscosity parameter controlling the dissipative resistance to changes in the nematic order parameter relative to a frame that  translates and rotates with the fluid, $\beta$ captures the reciprocal drag between fluid shear and changes in nematic order, and $\gamma>0$ is a friction parameter with a substrate.
	The entropy production inequality requires non-negativity and convexity of the dissipation potential, which is satisfied whenever
	\begin{equation}
		2\eta \eta_{\text{rot}}-\beta^2\ge0.
		\label{2nd_law}
	\end{equation}
See \ref{inequality} for a related derivation of this condition. We note that there is no thermodynamic restriction on the sign of $\beta$. Physically, however, the natural notion that filaments align along the direction of stretching is achieved by $\beta<0$ \cite{salbreux2009}.
	
	We define power input by
	\begin{equation}
		p(\bm{d},\widehat{\bm{q}};{\bm{q}}) =\lambda \bm{q}:\bm{d}^{\rm dev} - \lambda_{\bigodot}  \bm{q} : \widehat{\bm{q}},
	\end{equation}
	which accounts for an anisotropic active tension along the nematic tensor with activity parameter $\lambda$ and a generalized active force driving further alignment with activity parameter $\lambda_{\bigodot}$.
	
	The constraint integral must include an additional term accounting for incompressibility of the form
	\begin{equation} 
		\mathcal{Q} = \ldots + 
		\int_A P ({\rm tr} \bm{d}) dA, \label{eq:constraint_}
	\end{equation}
	where $P$ is a 2D pressure (with units of surface tension) acting as a Lagrange multiplier. 
	
	Particularizing the equations in the previous section, we can write balance of generalized force conjugate to changes in nematic order, i.e.~Eq.~(\ref{eq:balance_q}), as
	\begin{equation}
		\label{bal_gen_force}
		\widehat{\bm{q}} = \frac{1}{\eta_{\text{rot}}}\bm{h} -  \widetilde{\beta}  \bm{d}^{\rm dev} + \frac{\lambda_{\bigodot}}{\eta_{\text{rot}}}\bm{q},
	\end{equation}
	where we have introduced $\widetilde{\beta}= \beta/{\eta_{\text{rot}}}$, and the molecular field as
	\begin{equation}
		\label{nem_field2}
		h_{ab} = -\frac{\delta \mathcal{F}}{\delta {q}_{ab}} = -\frac{\partial f}{\partial {q}_{ab}} + \nabla_c  \frac{\partial f}{\partial \nabla_c {q}_{ab}}. 
			\end{equation}
	Balance of linear momentum becomes
	\begin{equation} 
		\label{bal_lin_mom}
		\gamma \bm{v} = \nabla\cdot\bm{\sigma},
	\end{equation}
where the stress tensor now accounts for the pressure  resulting from Eq.~(\ref{eq:constraint_}) and takes the form
	\begin{eqnarray}
		\label{stress_stand}
		{\sigma}_{ab}  =  -\frac{\partial  f}{\partial \nabla_b q_{dc}} \nabla_a q_{dc}  +  q_{ac} h_{cb} - q_{bc} h_{ca} + 2\eta {d}^{\rm dev}_{ab} + \beta \widehat{{q}}_{ab} + \lambda {q}_{ab} - P \delta_{ab}. 
	\end{eqnarray}
\mab{A direct calculations shows that the hydraulic pressure defined as $P_{\rm hydr} = - {\rm tr} \bm{\sigma}/2$ is related to the Lagrange multiplier enforcing incompressibility by the relation $P_{\rm hydr} = P + \frac{1}{2} {\partial  f}/{\partial \nabla_a q_{dc}} \nabla_a q_{dc}$.}
	
	In the literature, the dissipative coupling between strain rate and changes in nematic order appears in the stress tensor as a negative constant times $\bm{h}$ \cite{de1993}, rather than as our term $\beta \widehat{\bm{q}}$. To recover this expression, we insert the expression for generalized force balance, Eq.~(\ref{bal_gen_force}), into Eq.~(\ref{stress_stand}), to obtain
	\begin{equation}
		\label{stress_stand_2}
		{\sigma}_{ab}  = -\frac{\partial  f}{\partial \nabla_b q_{dc}} \nabla_a q_{dc}  +  q_{ac} h_{cb} - q_{bc} h_{ca} + 2\widetilde{\eta} {d}^{\rm dev}_{ab} + \widetilde{\beta} h_{ab} + \widetilde{\lambda} {q}_{ab} - P \delta_{ab},
	\end{equation}
	with effective shear viscosity $\widetilde{\eta} = \eta - \beta^2/(2\eta_{\text{rot}})$, which is positive according to Eq.~(\ref{2nd_law}), effective dissipative coupling coefficient $\widetilde{\beta}= \beta/{\eta_{\text{rot}}}$ and effective activity parameter $\widetilde{\lambda}= \lambda +  \lambda_{\bigodot} \beta/{\eta_{\text{rot}}}$. \mab{With this manipulation, it is readily verified that  balance of generalized nematic force in Eq.~(\ref{bal_gen_force}) and that the stress tensor in Eq.~(\ref{stress_stand_2}) involved in balance of linear momentum agree with Eqs.~(76,79) in \cite{julicher2018}, including the reciprocity relation according to which the same coefficient $\widetilde{\beta}$ appears in each of these equations.}
	
	We have thus shown that the natural expression of the stress tensor according to our theory,  Eq.~(\ref{stress_stand}), is equivalent to the  more conventional one for an incompressible model,  Eq.~(\ref{stress_stand_2}), provided that material parameters are reinterpreted. From a physical point of view, the interpretation of $\widetilde{\beta} h_{ab}$ in Eq.~(\ref{stress_stand_2}) as a dissipative term is somewhat indirect because $h_{ab}$ as defined in Eq.~(\ref{nem_field2}) depends only on the state of the system, and not on its rate-of-change. Instead, the interpretation of the dissipative stress induced by changes of nematic order relative to the fluid motion, $\beta \widehat{\bm{q}}$, is more natural. 	
	
\subsection{\mab{Variational derivation of a second model for active nematic fluids}}
\label{form_BE}

\subsubsection*{\mab{Governing equations based on a bracket formalism.}} 

\mab{We recover next another widely used model of active nemato-hydrodynamics \cite{marenduzzo2007,giomi2014,giomi2015,zhang2020,pearce2020,Axel_Voigt,metselaar2019,Santhosh2020,D2SM00988A} in the low Reynolds limit. With our notation and in 2D, the partial differential equation governing the evolution of the nematic field in this model takes the form \cite{Santhosh2020}
\begin{equation}
\partial_t q_{ab} + v_c \nabla_c q_{ab} - \mathcal{W}_{ab} = \Gamma h_{ab},
\label{nem_BE}
\end{equation}
with
\begin{equation}
\bm{\mathcal{W}} = \left(\alpha \bm{d} + \bm{w}\right)\left(\bm{q} + \frac{1}{2}\bm{I}\right) + \left(\bm{q} + \frac{1}{2}\bm{I}\right)\left(\alpha \bm{d} - \bm{w}\right) - 2\alpha (\bm{q} : \nabla\bm{v})\left(\bm{q} + \frac{1}{2}\bm{I}\right),\label{W_BE}
\end{equation}
where $\alpha$ is called the tumbling parameter. The stress tensor appearing in the equation of balance of linear momentum includes pressure, viscous, active, and passive-nematic components, and takes the form
\begin{eqnarray}
\bm{\sigma} & = & - P \bm{I} + 2\eta \bm{d}^{\rm dev} + \lambda \bm{q} - \frac{\partial f}{\partial \nabla \bm{q}}:\nabla \bm{q} + \bm{q}\bm{h} - \bm{h}\bm{q} \nonumber\\
& &  + 2\alpha(\bm{q}:\bm{h}) \left(\bm{q} + \frac{1}{2}\bm{I}\right) - \alpha \bm{h}\left(\bm{q} + \frac{1}{2}\bm{I}\right) - \alpha \left(\bm{q} + \frac{1}{2}\bm{I}\right)\bm{h},\label{stress_BE}
\end{eqnarray}
where in the fourth term the double contraction is with respect to the indices of the nematic tensor. If $f$ is quadratic in $\nabla \bm{q}$, this term takes the form $-\kappa \nabla_a q_{cd}\nabla_b q_{cd}$.}

\mab{By direct comparison with the previous section, we readily identify all the terms in stress tensor appearing in the first line of Eq.~(\ref{stress_BE}) as emerging from the incompressibility functional in Eq.~(\ref{eq:constraint_}), from the term $\eta \vert \bm{d}^{\rm dev} \vert^2$ in the dissipation potential, from the power input functional 
	\begin{equation}
		p(\bm{d};{\bm{q}}) =\lambda \bm{q}:\bm{d}^{\rm dev},
		\label{powpot_BE}
	\end{equation}
and from the nematic free energy density $f(\bm{q},\nabla \bm{q})$. We can further identify the rotational diffusivity as the inverse of the nematic viscosity, $\Gamma=1/\eta_{\rm rot}$. It is clear that the crux of the matter to frame this model in the variational formalism is to identify the suitable dissipation potential. }

\mab{As a preliminary to our variational derivation, we re-elaborate Eqs.~(\ref{nem_BE},\ref{W_BE}). Recalling the definition of the Jaumann derivative in Eq.~(\ref{eq:Jaumann}) and using the fact that the flow is incompressible, these equations can be directly rearranged as 
\begin{equation}
\label{eq:bal_gen_force_be}
\widehat{\bm{q}}-\alpha\bm{l} = \Gamma\bm{h},
\end{equation}
where we have introduced the symmetric and traceless tensor
	\begin{equation}
		\label{eq:def_l}
		\bm{l} = \bm{d}^{\rm dev} \left(\bm{q} + \frac{1}{2}\bm{I}\right) + \left(\bm{q} + \frac{1}{2}\bm{I}\right) \bm{d}^{\rm dev} - 2 (\bm{d}^{\rm dev}:\bm{q}) \left(\bm{q} + \frac{1}{2}\bm{I}\right).
\end{equation}
}

\subsubsection*{\mab{Heuristic motivation of the dissipation potential.}}

\mab{To identify the variational structure, we note that while Onsager's variational formalism is naturally formulated in terms of the dissipation potential, denoted by $\mathcal{D}(V;X)$ in the abstract formulation in \ref{Ons_Form}, bracket formalisms are naturally expressed in terms of the dual dissipation potential $\mathcal{D}^*$, related to $\mathcal{D}$ by a Legendre transform \cite{Kraaij2020}
\begin{equation}
\mathcal{D}(V;X) = \sup_{\xi}\left[\xi\cdot V -  \mathcal{D}^*(\xi;X) \right],
\end{equation}
where $\xi$ is a generalized force power-conjugate to the generalized velocity $V$. Equation (\ref{eq:bal_gen_force_be}) suggests viewing $\widehat{\bm{q}}-\alpha\bm{l}$ as a generalized velocity and $\bm{h}$ as a generalized force. The simplest form for the dual dissipation potential density for the nematic part then takes the quadratic form 
\begin{equation}
{d}^*_{\rm nem} (\bm{h}) = \frac{\Gamma}{2}\vert \bm{h}\vert^2, 
\end{equation}
and hence 
\begin{equation}
{d}_{\rm nem}  = \sup_{\bm{h}}\left[\bm{h}:\left( \widehat{\bm{q}}-\alpha\bm{l}\right)-  \frac{\Gamma}{2}\vert \bm{h}\vert^2\right]. 
\end{equation}
The stationarity condition of this optimization problem is precisely Eq.~(\ref{eq:bal_gen_force_be}), which plugged in this expression results in 
\begin{equation}
{d}_{\rm nem}  = \frac{\eta_{\rm rot}}{2}|\widehat{\bm{q}}-\alpha\bm{l}|^2,
\label{diss_nem_BE}
\end{equation}
in the spirit of the dissipation potential postulated in \cite{zhang2020}.}

\subsubsection*{\mab{Systematic variational derivation.}} 

\mab{Based on the heuristic argument above, we postulate a dissipation potential of the form
\begin{equation}
d(\bm{d},\widehat{\bm{q}}; \bm{q}) = \eta \vert \bm{d}^{\rm dev}\vert^2 
+ \frac{\eta_{\rm rot}}{2}  |\widehat{\bm{q}}-\alpha\bm{l}|^2.
\label{diss_pot_BE}
\end{equation}
Forming the Rayleighian and making it stationary with respect to $\widehat{\bm{q}}$, we immediately recover Eq.~(\ref{eq:bal_gen_force_be}), or equivalently Eq.~(\ref{nem_BE}). Particularizing the general form of the stress tensor in Eq.~\eqref{total_stress} taking into account the incompressibility contraint as in Eq.~\eqref{eq:constraint_}, we obtain
	\begin{equation}
		\label{eq:stress_be}
		\sigma_{ab} = -\frac{\partial  f}{\partial \nabla_b q_{cd}} \nabla_a q_{cd}  +  q_{ac} h_{cb} - h_{ac}q_{cb}  -   h_{cd} \frac{\partial l_{cd}}{\partial d^{\rm dev}_{ab}} + \lambda q_{ab} - P \delta_{ab} + 2\eta d_{ab}^{\rm dev}
	\end{equation}
	where we have used Eq.~\eqref{eq:bal_gen_force_be}. The derivative of $\bm{l}$ with respect to $\bm{d}$ can be computed from Eq.~\eqref{eq:def_l}
	\begin{equation}
		\frac{\partial l_{cd}}{\partial d^{\rm dev}_{ab}} =  \left[\delta_{ac} \left(q_{bd}+\frac{1}{2}\delta_{bd}\right) + \left(q_{ac} + \frac{1}{2}\delta_{ac}\right) \delta_{bd} - (2q_{ab}+\delta_{ab}) q_{cd}\right].
	\end{equation}
Plugging this expression in Eq.~\eqref{eq:stress_be}, we recover Eq.~(\ref{stress_BE}). }

\subsection{\mab{Discussion}} 

\mab{We have shown in Section \ref{form_BE} that the active version of the Beris-Edwards liquid crystal model in the low Reynolds limit minimizes a Rayleighian functional combining the nematic free energy, the dissipation potential in Eq.~(\ref{diss_pot_BE}) and the power potential in Eq.~(\ref{powpot_BE}) subject to the incompressibility constraint. Comparison with the derivation in Section \ref{form_IT} shows that the key difference between the two incompressible liquid crystal models is the dissipation potential. Our heuristic motivation of the Beris-Edwards model suggests that this model is natural from the point of view of a dual dissipation potential in terms of the molecular field. Instead, when modeling irreversible processes with a dissipation potential that depends on the rate-of-change of the system, the choice in Eq.~(\ref{diss_inc}) admits a clearer physical interpretation and is compatible with a rational constructive modeling approach. Instead, the dissipation potential in Eq.~(\ref{diss_pot_BE}) is quadratic in a quantity that does not admit a clear physical interpretation, and if expanded includes many more terms that cannot be easily ascribed to specific physical mechanisms. In turn, these terms lead to more complex Euler-Lagrange equations. 
%Our derivation also suggests that the Beris-Edwards model results from the simplest quadratic dual dissipation potential in terms of the molecular field. %We have not identified here the bracket structure of the model in Section \ref{form_IT}, but presumably its dissipative bracket and dual dissipation potential is more involved. 
Our derivations highlight that each of these models can be understood as resulting from natural modeling choices within a given formalism of irreversible thermodynamics, and therefore their suitability to understand specific physical systems should be assessed on their own merits in relation with experiments or microscopic models.}

\section{Variational derivation of a compressible density-dependent active nematic gel} \label{sec_3}
	
	Having derived the generic equations for a compressible active nematic gel in Section \ref{sec_2}, here we make specific choices for free-energy, dissipation and power input to derive the governing equations for a density-dependent active nematic gel. 
	
	For the free-energy density, we assume a Landau expansion
	\begin{equation} 
		\label{eq:landau}
		f(\bm{q},\nabla\bm{q}) = \frac{1}{2}a S^2 + \frac{1}{8}b S^4 + \frac{1}{2} L \left|\nabla \bm{q}\right|^2,
	\end{equation}
	where $L>0$ is the Frank constant penalizing gradients of orientation and $a$ and $b>0$ are susceptibility parameters. For $a>0$, the susceptibility parameters penalize deviations from the isotropic state given by $S=0$. For $a<0$, the susceptibility parameters penalize deviations from anisotropic states with $S= \sqrt{-2a/b}$. 
	
	For the dissipation potential, we adapt the form of $d$ in Eq.~(\ref{diss_inc}) to a 2D compressible thin layer. To motivate our functional form, we assume that this thin layer of gel is 3D incompressible with uniform and constant volumetric density $\rho^{3D}$ and thickness $h$; hence the areal density is $\rho = \rho^{3D} h$. The 3D rate of deformation tensor $\bm{D}$
	is block-diagonal with blocks $\bm{d}$ and the out-of-plane component $D_{33}$. Since ${\rm tr}\bm{D}=0$, it follows that $D_{33} = -{\rm tr}\bm{d}$ and hence $|\bm{D}|^2 = |\bm{d}|^2 + \left(\text{tr}\bm{d}\right)^2$  \cite{salbreux2009}. As a result, the dissipation potential in the gel can be expressed as $\int_V \eta^{3D} |\bm{D}|^2 dV = \int_A \eta^{3D} |\bm{D}|^2 h dA= \int_A (\eta^{3D}/\rho^{3D}) |\bm{D}|^2 \rho dA =  \int_A \eta(|\bm{d}|^2 + \left(\text{tr}\bm{d}\right)^2)\rho dA$ with $\eta =\eta^{3D}/\rho^{3D}$. Hence, we consider the dissipation density
	\begin{equation}
		\label{eq:diss_density_example}
		d(\bm{v},\bm{d},\widehat{\bm{q}}) =  \eta \left[|\bm{d}|^2 + \left(\text{tr}\bm{d}\right)^2\right] +\frac{\eta_{\text{rot}}}{2}  \left|\widehat{\bm{q}}\right|^2+ \beta  \bm{d}^{\rm dev}:\widehat{\bm{q}}  + \frac{\gamma}{2} \left|\bm{v}\right|^2.
	\end{equation}
	Clearly, Eq.~(\ref{eq:diss_density_example}) implies that $\mathcal{D}\left[\bm{0},\bm{0},\bm{0}\right]=0$. The condition
	\begin{equation}
		2\eta \eta_{\text{rot}} - \beta^2 \ge 0,
		\label{diss_ineq}
	\end{equation}
	further guarantees non-negativity and convexity of the dissipation potential, \ref{inequality}. Hence, Eq.~(\ref{diss_ineq}) ensures non-negative entropy production. \mab{We note that to formulate a density-dependent compressible model of the Beris-Edwards type, it would be sufficient to replace the second and third terms in Eq.~(\ref{eq:diss_density_example}) by the expression in Eq.~(\ref{diss_nem_BE}).}
	
	We consider the following power input density generated by out-of-equilibrium microscopic processes 
	\begin{eqnarray} 
		p(\bm{d},\widehat{\bm{q}};{\bm{q}})  & =  \lambda \textup{tr}\bm{d} + \lambda_{\rm aniso} \bm{q}:\bm{d} - (\lambda_{\bigodot}' + \rho \lambda_{\bigodot})  \bm{q} : \widehat{\bm{q}}  \nonumber \\ 
		& =   \lambda \left(\bm{I} + \kappa \bm{q}\right) : \bm{d} - \rho \lambda_{\bigodot}  \bm{q} : \widehat{\bm{q}}, 
		\label{eq:pow_density_example}
	\end{eqnarray}
	where the first term in the first line is the power of an isotropic active tension, which now makes sense because of compressibility, the second term is the power of an anisotropic active tension along the  nematic tensor, and the third term is the power of an active generalized force conjugate to changes in nematic order. In contrast to the previous section, here we expand the corresponding activity parameter up to linear order in density. The constant term $\lambda_{\bigodot}'$ has a formally equivalent effect in the governing equations as the first term in Eq.~(\ref{eq:landau}), and hence can be subsumed in susceptibility parameter $a$.  For this reason we consider $\lambda_{\bigodot}'= 0$ in the second line, where  we group active tensions in a single term by defining the tension anisotropy parameter $\kappa=\lambda_{\rm aniso}/\lambda$. For $\lambda_{\bigodot}>0$, nematic activity tends to further increase alignment.

	With the free-energy, dissipation and power-input functions in Eqs.~(\ref{eq:landau},\ref{eq:diss_density_example},\ref{eq:pow_density_example}), Onsager's variational formalism developed in Section~\ref{sec:Onsager} yields the following generalized force balance equation
	\begin{equation}  \label{first_govern}
		\eta_{\text{rot}} \widehat{\bm{q}} + \beta \bm{d}^{\rm dev} + (2a + b S^2)  \bm{q} - L \left(\Delta \bm{q} +   \nabla\bm{q} \cdot \frac{\nabla \rho}{\rho} \right) - \rho\lambda_{\bigodot} \bm{q} = \bm{0}.
	\end{equation}
	This equation shows that $2a-\rho \lambda_{\bigodot}$ can be interpreted as an effective density-dependent susceptibility coefficient, which if negative, triggers spontaneous ordering. 
	
	Balance of linear momentum takes the form
	\begin{equation}
		\label{eq:balance_forces_linear}
		\nabla\cdot\bm{\sigma} = \rho \gamma \bm{v},
	\end{equation}
	where the stress tensor is the sum of its symmetric component
	\begin{equation}
		\sigma^{\rm s}_{ab} = \rho \left[2\eta  (d_{ab}+d_{cc} \delta_{ab}) + \beta  \widehat{q}_{ab}  + \lambda \left(\delta_{ab} + \kappa q_{ab}\right) -L\nabla_a q_{cd} \nabla_b q_{cd}\right],
	\end{equation}
	and its antisymmetric component 
	\begin{equation}
		\sigma^{\rm a}_{ab} = L \left[\nabla_c \rho \left( q_{ad} \nabla_c q_{db}  - q_{bd} \nabla_c q_{da}  \right)   + \rho  \left(q_{ae}\Delta q_{be}  -q_{be}  \Delta q_{ae}  \right) \right].
	\end{equation}
	For the boundary conditions, we either consider examples where $\bm{t}$, $\bm{\Gamma}$ and $\bm{L}$ are zero, leading to homogeneous Neumann boundary conditions, or examples where $\bm{v}$ and $\widehat{\bm{q}}$ are fixed, leading to Dirichlet boundary conditions. For the right-hand side of Eq.~(\ref{eq:balance_mass}), we consider a polymerization rate $k_p$, a depolymerization rate proportional to $\rho$ and given by $-k_d \rho$, and Fickian diffusion with diffusivity $D$,
	\begin{equation} \label{last_govern}
		\dot{\rho} + \rho {\rm tr}~\bm{d} = k_p - k_d \rho + D \Delta \rho.
	\end{equation}
	
	We end this section by briefly discussing how this model can lead to patterns of nematic order. The conventional mechanism for nematic ordering is driven by the free energy $f$, e.g.~as a result of the excluded volume effects in the mixing entropy of elongated particles \cite{de1993}. The model presented here can account for this mechanism by considering $a<0$, as in the numerical study in Section \ref{ex2}. \mab{However, nematic order can also arise from an initially isotropic systems ($a>0$) as a result of activity, either imposed to be heterogeneous \cite{salbreux2009} or, more interestingly, emerging from a symmetry-breaking instability leading to nematic patterns, as shown in \cite{Santhosh2020}  with an incompressible active nematic model and in \cite{Mirza_2024} with the present model. We note that the physical nature of the mechanisms of active nematic pattern formation in these two references is very different. In \cite{Mirza_2024} it is the result of the coupling between nematodynamics and an advective instability  typical of actomyosin gels, according to which self-reinforcing convergent flows produce velocity gradients and density accumulation  that locally increase order as a result of terms $\beta \bm{d}^{\rm dev}$ and $- \rho\lambda_{\bigodot} \bm{q}$ in Eq.~(\ref{first_govern}).}

	\section{Computational studies of compressible active nematodynamics}\label{sec_5}

	\subsection{Cell wound healing}
	\label{ex1}
	
	The dynamical assembly of a contractile ring is essential to close holes and tears formed in the cortex of the Xenopus egg, as systematically examined with laser ablation experiments \cite{benink2000,mandato2001}. Ablation triggers a localized stimulus increasing myosin-II activity at the edge of the wound. A more contractile region in the actomyosin gel produces a gradient in active tension driving long-range cortical flows. The interplay between actin flow and enhanced myosin activity leads to the assembly of a ring made of a dense network of interconnected F-actin bundles, myosin-II and other actin-binding proteins. Inside the contractile ring, the network is aligned parallel to the boundary of the wound. This contractile ring leads to wound closure by the purse-string mechanism \cite{Alice2008}. \mab{We motivate our study in experiments of wound healing at the subcellular scale \cite{benink2000,mandato2001} because this situation was analyzed previously with a simplified theoretical model related to ours \cite{salbreux2009}, but similar processes take place at supracellular scales \cite{10.1083/jcb.201804048}.}

	\begin{figure}[t]
		\centering
		\includegraphics[width=0.80\textwidth]{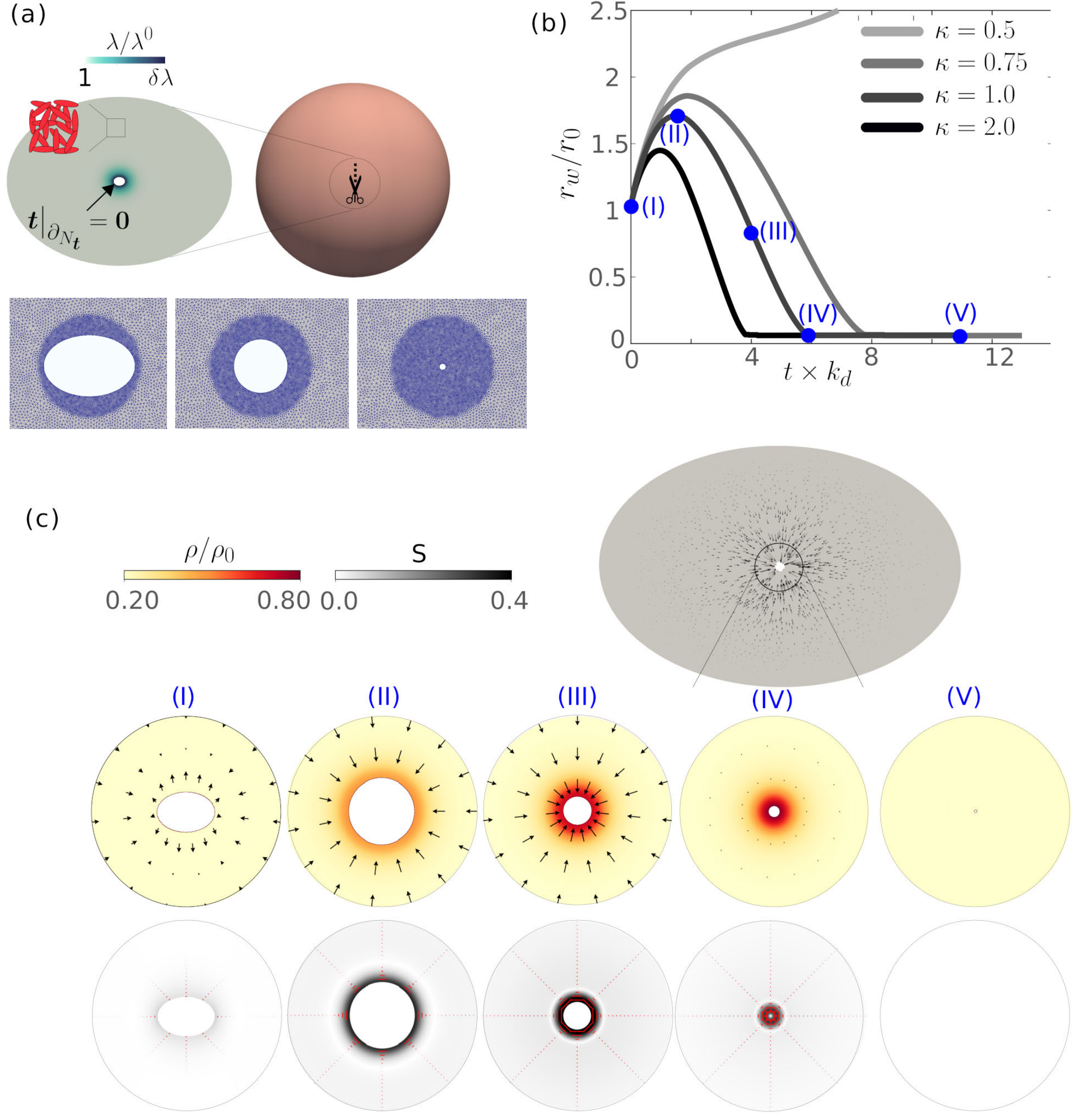}
		\caption{\label{sec_1_chap_3_fig_1} \textbf{Simulation of the process of wound healing}. (a) Schematic of cell wound healing setup \mab{and of the mesh refinement strategy to deal with the moving domain problem}. (b) Normalized wound size, where $r_w$ is the characteristic wound size and $r_0$ the minor axis of the initial wound, as a function of dimensionless time for different values of active tension anisotropy $\kappa$. (c) Snapshots of flow and density fields (top) and nematic field (bottom, segments denote nematic direction and colormap indicates $S$) close to the wound. See also Movie~1.}
	\end{figure}
	
We examine the self-organization of the contractile ring using the active nematic gel theory presented in Section \ref{sec_3} and study conditions leading to wound closure. The model setup is  illustrated in Fig.~\ref{sec_1_chap_3_fig_1}(a). We make the following assumptions. We ignore the curvature of the cell and consider a  planar patch of actin cytoskeleton. The characteristic size of the domain $ \ell_0$ is much larger than any inherent length-scales of the model such as the hydrodynamic length scale $\ell_s=\sqrt{\eta/\gamma}$ or the nematic correlation length scale $\ell_p=\sqrt{L/|2a-\rho_0\lambda_{\bigodot}|}$. This assumption implies that at distances greater than $\ell_s$ and $\ell_p$, the effect of the ablated region on the system is negligible. We represent the \mab{initial} ablated region as an ellipse with aspect ratio $1.5$ and  minor axis $r_0$. We denote the characteristic wound size with $r_w(t)$ calculated as the minimum distance from the edge of the wound to its centroid. To reflect an enhanced activity around the ablated region, we set the parameters that control active tension to
	\begin{eqnarray} 
		\lambda(\bm{x},t)=  \lambda^{0} \left( 1+  \delta \lambda\text{e}^{-r(\bm{x},t)/w}  \right), \;\;\;\;\mbox{and}\;\;\;\;
		\lambda_{\bigodot}(\bm{x},t)=  \lambda^{0}_{\bigodot} \left( 1+   \delta \lambda\text{e}^{-r(\bm{x},t)/w}  \right), \label{eq:over_activity_1}
	\end{eqnarray}
	where $r(\bm{x},t)$ is the distance between point $\bm{x}$ and its closest point projection on the boundary of the wound at time $t$.  In Eq.~(\ref{eq:over_activity_1}), $\lambda^0$ and  $\lambda_{\bigodot}^0$ are the base activity parameters and  $\delta \lambda$ sets the amplitude of the enhanced activity, which decays with the distance to the wound edge. We consider the width of the over-activity region $w$ to be larger than $\ell_p$ and smaller than $\ell_s$. \mab{We assume homogeneous Neumann boundary conditions, i.e.~all boundary tractions  ($\bm{t}$, $\bm{\Gamma}$ and $\bm{L}$) vanish at the wound edge. This means in particular that we do not impose any anchoring boundary condition.}
	
	\mab{The initial conditions are those of a uniform, quiescent and isotropic gel in its steady-state. We choose the parameters so that the effective susceptibility $2a -  \rho_0\lambda_{\bigodot}$ is positive, and hence any nematic ordering is of active origin}. All material parameters are given in Table~\ref{modal_parameters_cell_healing}. To track changes of domain shape during wound closure, we adopt an updated  Lagrangian approach; at each time-step, we update the nodes of the mesh with the velocity field. To maintain the mesh quality during this Lagrangian flow, after updating the nodal coordinates of the mesh, we perform reparametrization of the mesh while keeping the boundary fixed if the local element distortion  exceeds a given threshold.  Then, the fields $\bm{q}$ and $\rho$ are projected in the reparametrized mesh by a least-squares procedure.

	We first examine the role of $\kappa$, characterizing the anisotropy of active tension, as shown in Fig.~\ref{sec_1_chap_3_fig_1}(b). To illustrate the process of wound healing according to our model, we first focus on the curve corresponding to $\kappa=1.0$ with five snapshots I to V showing wound shape, density, velocity and nematic order, Fig.~\ref{sec_1_chap_3_fig_1}(c).  At time-point I, we impose a local increase of activity following Eq.~(\ref{eq:over_activity_1}). At this instant and due to tension in the active gel, the wound edge retracts away from the center increasing the size of the wound. At the same time, a centripetal actin flow driven by the gradient in active tension develops in a region of size commensurate to the hydrodynamic length. This convergent flow locally densifies the gel close to the edge of the wound, which further reinforces contractility and  flow towards the edge.  The centripetal flow rapidly decreases at the wound edge (it actually changes sign between instants I and II), generating a strong velocity gradient. Due to the flow-alignment effect associated to $\beta<0$, this velocity gradient increases nematic order near the edge parallel to it, whereas far away nematic order is weak with alignment perpendicular to the wound edge. The localized density and nematic order in the wound edge mobilize the generalized active nematic force $-\rho\lambda_{\bigodot}\bm{q}$ further driving order. In summary, local edge overactivity along with the traction-free boundary conditions lead to the self-assembly of a dense contractile bundle with high alignment parallel to the edge. \mab{Interestingly, in our compressible and contractile system, activity drives parallel anchoring at the boundary, opposite to the homeotropic anchoring induced by activity in other incompressible and contractile active nematic models \cite{doi:10.1098/rsta.2020.0394}.} Since here $\kappa >0$, contraction is larger along the nematic direction and hence this ring is highly contractile, creating a Laplace-like force in the curved wound edge, which tends to make it circular and overcomes cortical surface tension to close the wound, Fig.~\ref{sec_1_chap_3_fig_1}(III,IV). As the wound closes, the architecture of the contractile ring is stabilized by the interplay of cytoskeletal self-enhancing flows, flow-induced alignment, active alignment, diffusion, and  turnover. This leads to a robust process of wound healing.  When the size of the wound is very small relative to all other length-scales of the problem, we consider that the wound has closed and hence set the overactivity signal $\delta \lambda=0$. As a consequence,  the self-reinforcing flows rapidly decrease and  the contractile ring disassembles as cytoskeletal density reduces due to turnover (V). Hence, following a largely self-organized process of wound healing, the cortex recovers homeostasis. See Movie 1 for an illustration. A parametric sweep for different values of $\kappa$ shows that $\kappa$ needs to be sufficiently high for wound closure, Fig.~\ref{sec_1_chap_3_fig_1}(b).

	\begin{figure}[t]
		\centering
		\includegraphics[width=0.75\textwidth]{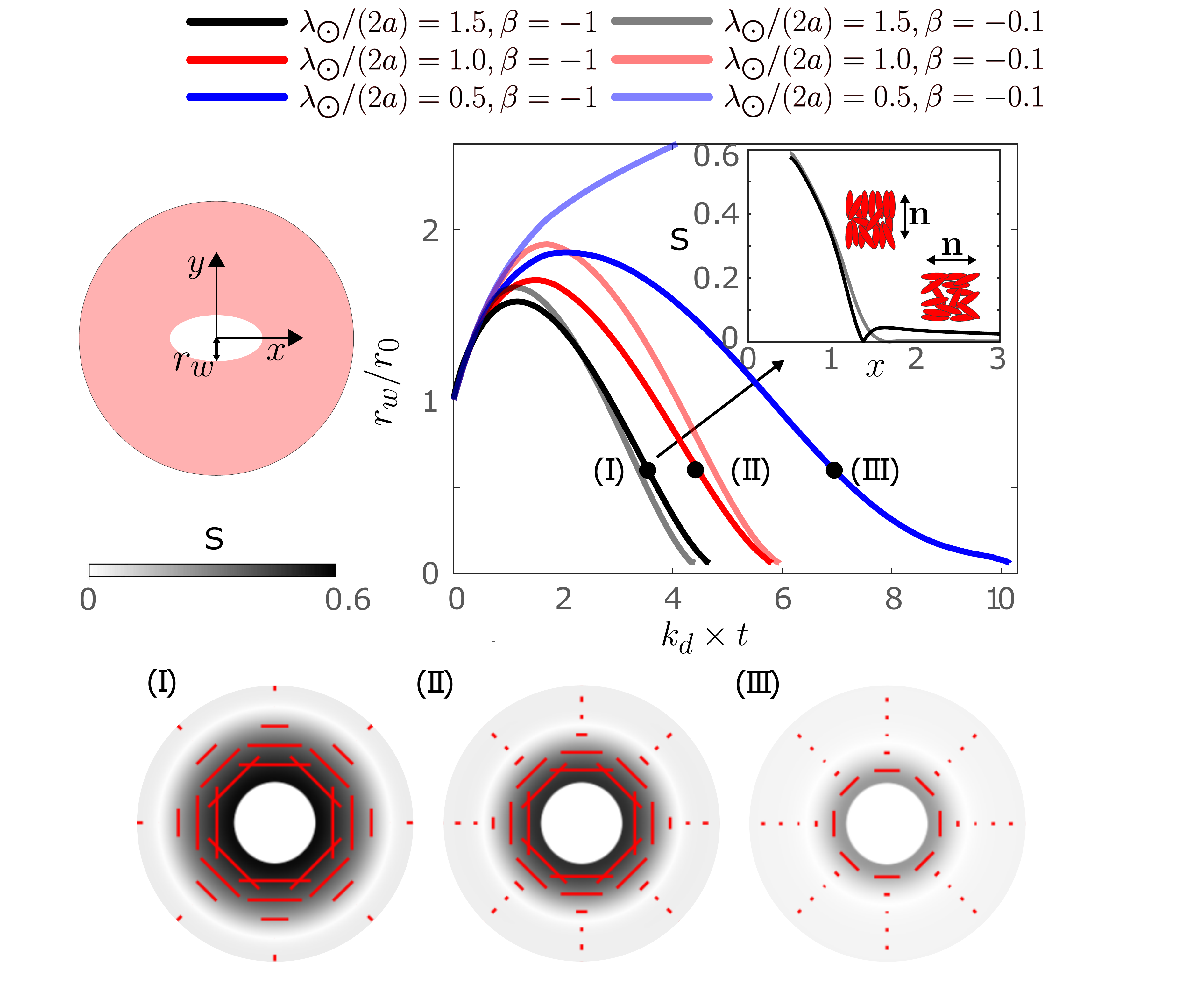}
		\caption{\label{sec_1_chap_3_fig_2}  \textbf{Effect of the flow-aligning parameter $\beta$ and nematic activity $\lambda_{\bigodot}$ on the healing dynamics}. Normalized wound size  $r_w$ as a function of time for different choices of parameters. The snapshots (I-III) show the nematic organization at a given wound size for different $\lambda_{\bigodot}$. The nematic field is represented by a color map for $S$ and by red segments, whose direction indicates the nematic orientation and whose size is proportional to $S$.}
	\end{figure}

	We then examined the effect of $\lambda_{\bigodot}$ and $\beta$ on wound healing,  Fig.~\ref{sec_1_chap_3_fig_2}. A higher value of $\lambda_{\bigodot}$ self-organizes a contractile ring with a higher nematic order on a shorter time scale. This leads to a quicker inhibition of  wound opening and of  reversal of  boundary motion. Contrasting with this strong effect, the flow-aligning parameter $\beta$ has a milder and more subtle effect. A higher $\beta$ promotes the fast formation of the contractile ring, but also aligns radially the network far away, see inset, counteracting the effect of the ring. Close to the threshold between wound closing and opening, changes in $\beta$ can have a dramatic effect on the dynamics, see blue curves.

	\mab{In summary, our model for a density-dependent and compresible active nematic fluid captures the dynamics of would healing observed in the actomyosin cytoskeleton of large cells, in a largely self-organized process. The model only introduces as an input signal an over-activity close to the rim of a wound. Downstream of this signal, compressive flows align the nematic field and trigger the assembly of a dense and contractile nematic cable that restores homeostasis. Compressibility and the density-dependent self-reinforcing flows and active alignment (depending $\lambda_{\bigodot}$) are essential to the formation of a dense and aligned structure (the ring) in a low-density and low-order background (the surrounding cortex), as further studied in \cite{Mirza_2024}. Therefore, the mechanism recapitulated by our model cannot be recapitulated by the most common models for active nematic fluids.}

	\subsection{Defects in a confined colony of spindle-shaped cells}
	\label{ex2}

	Having examined a situation where nematic order has an active origin linked to self-reinforcing flows, we turn now to a more conventional situation in which \mab{ordering is driven by the close packing of elongated objects ($a<0$), and study the effect of confinement, activity and compressibility on the dynamics of the system. Diverse dense active nematic systems have been subjected to confinement in disk-like domains, including kinesin-microtubule gels \cite{norton2018} or elongated spindle-shaped contractile cells such as myoblasts or fibroblasts \cite{duclos2014,guillamat2020}, both of which exhibit long-range nematic order. Cells confined in these circular domains tend to align parallel or perpendicular to the boundary \cite{guillamat2020}.
Because of this boundary alignment and the spontaneous tendency to nematic ordering, the net charge of the topological defects in the colony is $+1$, as required by the Poincar\'{e}-Hopf theorem. Because nematic order controls the anisotropy of active stresses in the cell monolayer, activity may lead to a variety of out-of-equilibrium behaviors including persistent flows or defect motion. }

\mab{The influence of confinement and activity in such situation was studied computationally in \cite{norton2018}, focusing on kinesin-microtubule systems and considering an extensile incompressible system. Here, we focus on a contractile system representative of some nematic dense cell colonies. Due to the cellular compressibility and to cell proliferation or extrusion, the incompressible flow assumption may not be pertinent to study cellular colonies. See for example the convergent flows around defects reported in \cite{blanch2018,guillamat2020}. To isolate the role of compressibility, we compare two limiting active nematic models where a uniform initial density remains constant during the dynamics. In a first limit, we consider an incompressible flow, so that the model in Section \ref{sec_3} reduces to the model in Section \ref{form_IT}. In a second limit, we consider fast turnover, so that Eq.~(\ref{last_govern}) is in a steady state and convergent/divergent flows are allowed. In both limits, the density field becomes a constant, leaving us with the coupling between nematic and velocity fields. See \cite{PhysRevE.106.054610} for a related compressible/incompressible comparison for an extensile system.}

%	Depending on the size of the geometrical confinement with respect to the characteristic lengths of the system, self-organized flows and motion of defects are either absent \cite{duclos2014}. \WAM{In \cite{norton2018}, spiral or turbulent-like patterns were observed in an extensile, circularly confined system, with behavior varying depending on the size of the confined domain. However, in this section, we focus on a contractile compressible active nematic system, which is relevant for reconstituted systems in flat two-dimensional \cite{duclos2014} and spherical three-dimensional surfaces \cite{litschel2021}. For completeness and physical relevance in microorganism systems \cite{ronning2022}, we also discuss the behavior of incompressible contractile active nematic systems.} \mab{[This paper deserves more comment. It is also incompressible. How does it compare with your new results? ]}  \WAM{Marino, later i highlight a bit how the behaavior of turbulence in contractile system is different hte extensile system}. 
	
	\begin{figure}[t]
		\centering
		\includegraphics[width=0.85\textwidth]{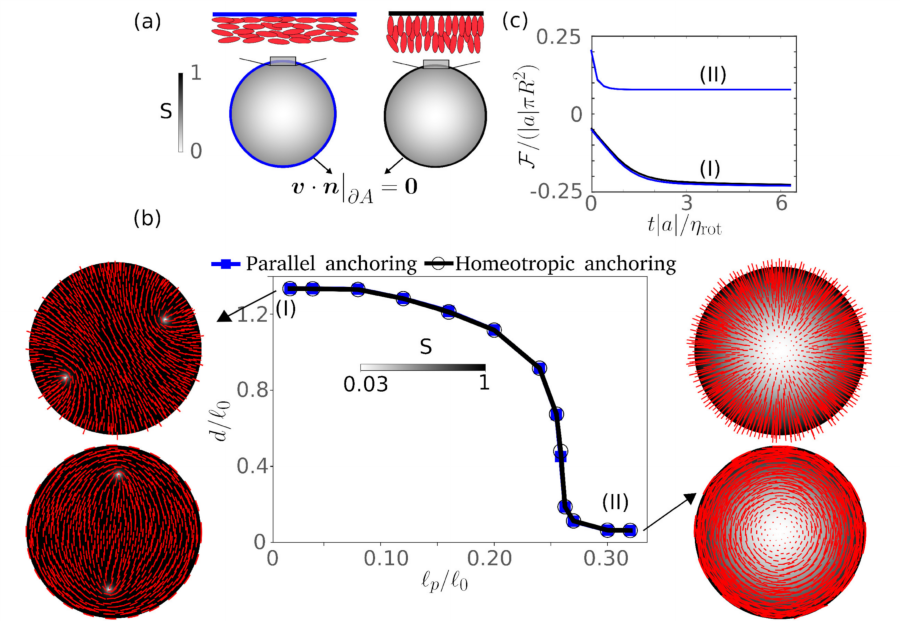}
		\caption{\label{sec_1_chap_3_fig_3} \textbf{Effect of nematic correlation length scale $\ell_p$ on the inter-defect distance $d$ in a passive nematic system}. (a) Illustration of nematic and velocity boundary conditions. (b) Inter-defect distance as a function of nematic correlation length scale $\ell_p$ for parallel and homeotropic anchoring conditions, along with selected nematic fields. (c) Monotonically decreasing time-evolution of the Landau free-energy for high and low $\ell_p$ and for both boundary conditions.}
	\end{figure}
	
  In order to model the propensity of cells towards mutual alignment, we set the susceptibility parameter such that the initial nematic order is close to $S_0=\sqrt{-2a/b}=1$. All model parameters are detailed in Table~\ref{modal_parameters_cell_colony}, \mab{and are identical for the compressible and the incompressible versions of the model}. We impose a boundary condition such that $S=1$ at the boundary and the director field is aligned either tangentially ($\bm{n}$ perpendicular to  $\bm{N}$) or perpendicularly ($\bm{n} = \bm{N}$) to the boundary, corresponding to parallel or homeotropic anchoring, see Fig.~\ref{sec_1_chap_3_fig_3}(a). We enforce that velocity normal to the boundary is zero (impermeable boundary) by introducing a penalty term in the Rayleighian given as  $\int_{\partial A} K \left|\bm{v} \cdot \bm{N}\right|^2 dl$, where $K$ is the penalty coefficient, but allow cells to slide tangentially to the boundary.
	
	We first examine the behavior of a passive nematic system for different ratios between the nematic correlation length  $\ell_p = \sqrt{L/\left(2|a|\right)}$ and the radius of the domain  $\ell_0$.  We start with a spatially correlated random initial condition for $\bm{q}$.  In agreement with previous results, we find that two $+1/2$ defects initially nucleate near the boundary, and then travel away from the wall into the bulk, reaching a quiescent steady-state \cite{giomi2014}. During this process, the free-energy decreases,  Fig.~\ref{sec_1_chap_3_fig_3}(c), and at steady-state velocities and rate of dissipation vanish. For systems with parallel (homeotropic) boundary conditions, the tips of the $+1/2$ defects point away from (towards) each other, Fig.~\ref{sec_1_chap_3_fig_3}(b). The size of the defect cores relative to system size is controlled by $\ell_p/\ell_0$. As this quantity increases, we expect the two defects to interact and possibly combine into a single $+1$ defect as observed in small-size cell colonies. To examine this, we track the distance between the two defects, $d$, as a function of \mab{the confinement parameter} $\ell_p/\ell_0$, finding that beyond a threshold, $d$ abruptly drops close to zero, with configurations resembling a vortex or an aster depending on boundary conditions, Fig.~\ref{sec_1_chap_3_fig_3}(b). We note, however, that in the absence of activity, $d$ stays finite, and hence the two defects do not strictly become a $+1$ defect, which can be understood in terms of a Coulomb-like repulsion \cite{vafa2020}. 
		
\begin{figure}[tb]
		\centering
		\includegraphics[width=0.77\textwidth]{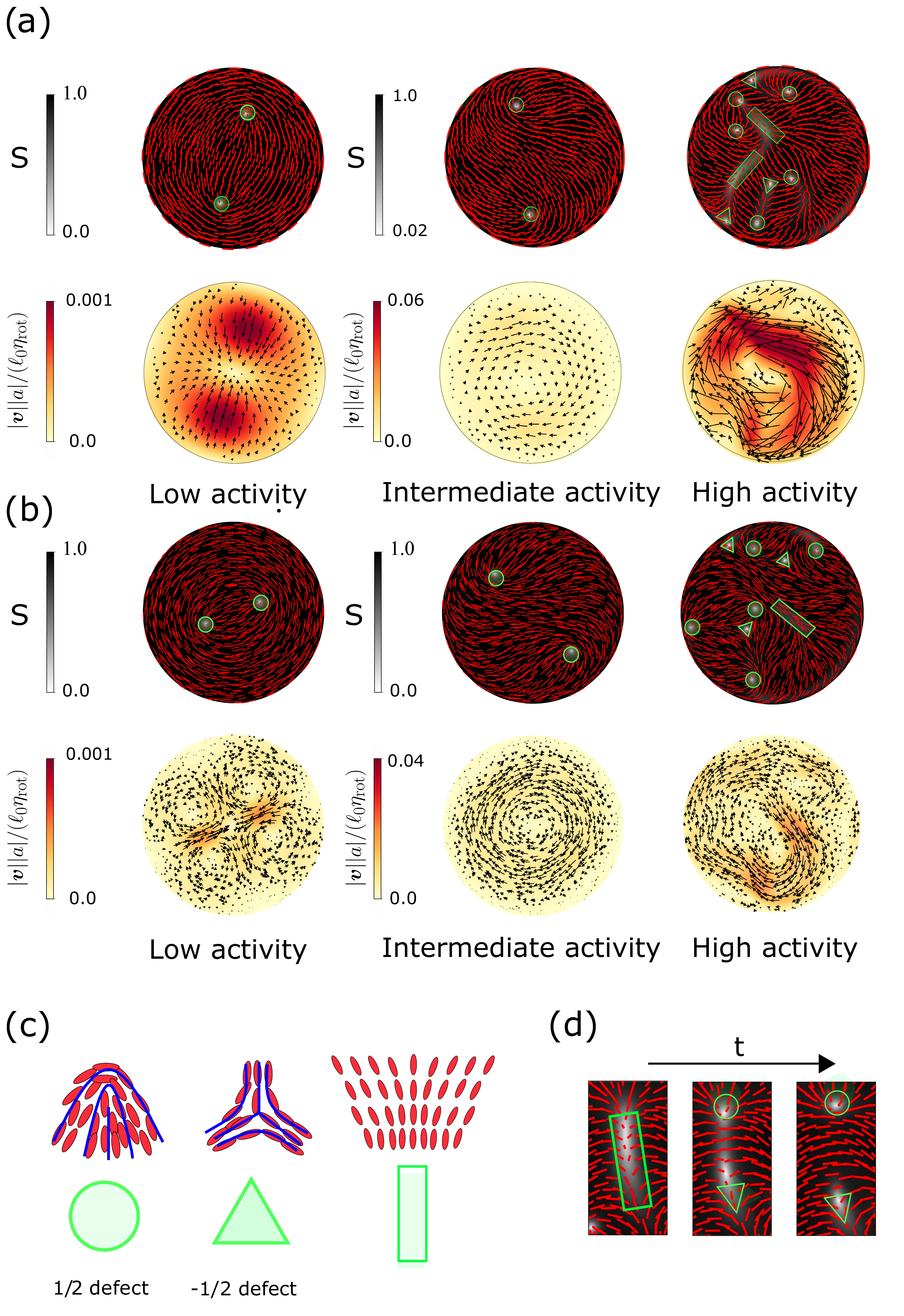}
		\caption{\label{sec_1_chap_3_fig_3.5}  \textbf{Effect of Activity on a Dense, Confined Colony of Contractile Cells.} \mab{Snapshots of nematic and velocity fields depending on activity for a compressible (a) and in incompressible (b) active nematic model. See Movies~2 and~3 for the dynamics of the compressible system and Movie~4 for the  incompressible system. (b) Schematic representation of $\pm \frac{1}{2}$ defects and splay bands. (c) Nucleation of $+1/2$ and $-1/2$ defects from splay bands in the high-activity regime.}}
	\end{figure}

We next explore the spatiotemporal behavior of defects in a contractile ($\lambda_{\rm aniso} > 0$) active nematic system by examining the  velocity and nematic fields as a function activity, measured by the active length scale $\ell_a = \sqrt{L/\lambda_{ \rm aniso}}$. We vary this inverse activity parameter at low \mab{confinement} $\ell_p/\ell_0 = 0.02$ and focus on parallel boundary conditions for the nematic field, see Fig.~\ref{sec_1_chap_3_fig_3.5}.  At low activity (large $\ell_a/\ell_0$), we observe that the contractile active  stress $\lambda_{\rm aniso} \bm{q}$ drives the system to new steady state with a smaller inter-defect distance \mab{for both the compressible and the incompressible models,  see left panels of Fig.~\ref{sec_1_chap_3_fig_3.5}(a, b) and Movies~2 and 4. More importantly, the steady state of the active system exhibits persistent flows in the nose to tail direction around defects \cite{doostmohammadi2018}, which push these defects by advection of nematic order, Eq.~(\ref{jaumann_detivative_def}). The passive nematic distribution is also distorted  by the flow-induced alignment term involving $\beta$. Not surprisingly, the steady-state flow patterns of the compressible and the incompressible models are starkly different; while the incompressible model leads to a recirculating field organized around four vortices as previously described \cite{norton2018,doostmohammadi2018}, the compressible model develops a convergent flow field in the axis of the defects. The larger magnitude of the velocity field in the compressible case can be explained by the smaller velocity gradients allowed by compressibility, leading to smaller viscous drag.}

	\mab{At intermediate activity, the non-equilibrium steady-state characterized by two quiescent $+1/2$ defects becomes unstable, and the two defects acquire a chiral configuration leading to persistent rotation of the pair of defects and a vortical flow field. This behavior is observed for both the compressible and the incompressible systems, see middle panels in Fig.~\ref{sec_1_chap_3_fig_3.5}(a,b) and Movies~2 and 4. This behavior is reminiscent of that reported in confined cell colonies in \cite{guillamat2020}. Further increasing activity leads to seemingly chaotic dynamics with multiple defects for both models, see right panels in Fig.~\ref{sec_1_chap_3_fig_3.5}(a,b) and Movies~3 and 4. In this regime, active stresses overcome the restoring effect of stresses resulting from the nematic energy, distorting the nematic field to form lines of splay-type disclination in the bulk and close to the boundary. These splay bands lines destabilize and split into pairs of $\pm 1/2$ defects as previously described \cite{ramaswamy2007}, see Fig.~\ref{sec_1_chap_3_fig_3.5}(c,d), which move and annihilate with defects of opposite charge. The persistent generation of disclination lines, their destabilization into point defects, and the motion and annihilation of these defects gives rise to a behavior akin to active turbulence. At any given time, the system exhibits more than two defects but the total topological charge is conserved to $+1$. }

\mab{In summary,  our simulations exhibit a diversity of dynamical regimes depending on activity. Because a gradient in nematic order in the vicinity of defects generates an active flow, defects become motile. At low activity, the pair of defects are driven together to a new steady-state and generate strikingly different flow fields depending on whether the system is compressible or incompressible. At intermediate activity, 
the two defects break symmetry and wobble closer to the boundary, or develop a persistent spiral motion accompanied by a vortical flow field (Movies 2 and 4). At high activity, we observe active turbulence characterized by persistent defect nucleation due to splay-type instabilities, motion and annihilation. Unlike the case of low activity, when defects are motile the qualitative features of the dynamics are similar for the compressible and  incompressible models. We end by mentioning that in our compressible and incompressible contractile systems, the motion of defects is opposite to that observed in extensile active nematic systems \cite{norton2018}. This difference arises because the hydrodynamic flow around $\pm 1/2$ defects is reversed. Moreover, defect nucleation in turbulent regimes is driven here by splay-type instabilities rather than bend-type instabilities characteristic of extensile  systems.}

	\section{Summary and outlook} \label{summary}
	
	\mab{We have proposed a general modeling framework for density-dependent active nemato-hydrodynamics in 2D. This framework relies on Onsager's variational formalism for irreversible thermodynamics, providing a simple and direct  procedure to develop thermodynamically consistent models of active and structured fluids in fully nonlinear regimes, as well as to develop finite element methods. We have shown that this formalism enables a clear and systematic derivation of otherwise complex governing equations coupling nematic order, gel velocity and density. }
	
\mab{	We have shown that two commonly used models for incompressible active nematic fluids can be naturally cast within the proposed variational framework, providing additional physical insight on their underlying structure and modeling assumptions. We have further developed a new density-dependent active nematic fluid gel. As shown elsewhere \cite{Mirza_2024}, this model actively self-organizes patterns of dense nematic structures starting from an isotropic and quiescent gel, which can explain the polymorphism and dynamics of the actomyosin cytoskeleton. We have developed a numerical finite element method to approximate this model and applied it to two studies of biological relevance, which highlight the role of compressibility of the active flows. In the first study, we have explored the role of the self-organization of the actin cytoskeleton during wound repair. In our simulations,  overactivity around the wound drives a self-reinforcing compressive flow of the actin gel, leading to the self-organization of a nematic bundle that efficiently constricts the wound, in close analogy with experimental observations. In this example, nematic oder arises due to activity, as self-reinforcing flows locally densify and orient nematic order. Such an autonomous mechanism for recovery of cortical homeostasis cannot be described by common incompressible models of active nematic fluids. In a second numerical study, we explore the self-organization of a dense and confined active nematic system. We consider two limits of the model, namely incompressibility and high turnover. Depending on the magnitude of the activity, the topological defects required by boundary conditions either reach a steady state, with very different flow patterns in the compressible and incompressible cases, or exhibit highly dynamical flows as well as active turbulence. }

\mab{	If suitably extended to curved and time-evolving surface domains, the framework presented here can help elucidate the interaction between nematodynamics and reshaping during morphogenesis from cellular to organism scales \cite{mayer2010,maroudas2021}. Future work may also exploit the optimality principle underlying active nematic models derived from Onsager's formalism to develop  variational physics-informed machine learning models that identify the nature of free-energy, dissipation and activity from experimental observations \cite{PhysRevLett.129.258001}.}

	\section*{Acknowledgments}
	The authors acknowledge the support of the European Research Council (CoG-681434) and the Spanish Ministry for Science, Innovation and Universities (project PID2022-142178NB-I00 funded by MICIU/AEI/ 10.13039/501100011033 and ERDF/EU). WM acknowledges the La Caixa Fellowship  and the European Union’s Horizon 2020 research and innovation program under the Marie Skłodowska-Curie action (GA 713637). MA acknowledges the Generalitat de Catalunya (ICREA Academia prize for excellence in research). IBEC and CIMNE are recipients of a Severo Ochoa Award of Excellence.
	
	\appendix
	
	\section{\mab{Abstract summary of Onsager's variational formalism}}
	\label{Ons_Form}
	
	We sketch here a minimal abstract formulation of the variational formalism. We denote the state variables of the system as $X(t)$, the system free energy as $\mathcal{F}(X)$, the process variables describing how the systems changes as $V$, a dissipation potential as $\mathcal{D}(V;X)$, and a potential for the external/active power input as $\mathcal{P}(V;X)$,  a linear operator in $V$ taking the abstract form $\mathcal{P}(V;X) = -F(X) V$ where $F(X)$ are external/active generalized forces. By ``$(V;X)$'' we emphasize that the main dependence is on $V$ but that there may be a parametric dependence on $X$. The state and process variables may include chemical, structural and mechanical fields. We also suppose that the process variables are constrained by $0 = \mathbb{C}(X) V$. We shall assume that all these potentials satisfy frame indifference and material symmetries, and that $\mathcal{D}$ is nonnegative, satisfies $\mathcal{D}(0,X) = 0$, is a differentiable and convex function of $V$, but need not be quadratic \cite{EDELEN1972481,PhysRevE.47.351,mielke2016generalization}. The free energy may be nonlinear and non-convex. 
	
	In general, the process variable $V$ may not be simply $\partial_t X$. For instance, in a model dependent on density $\rho$ advected by a flow with velocity field $\bm{v}$, the continuity equation relates the rate of change of state variable $\partial_t \rho$ with the process variable $\bm{v}$. As noted by \cite{Otto2001,peletier2014variational}, $V$ often contains redundant information to describe $\partial_t{X}$, which is however required to properly model dissipation. Indeed, in the example above $\partial_t \rho$ is a scalar field but $\bm{v}$ is a vector field. We formalize the relation between $\partial_t X$ and $V$ through a linear process operator 
	\begin{equation}
		\label{general_process}
		\partial_t{X}=P(X)V.
	\end{equation} 
	The rate of change of the free energy follows from the chain rule and Eq.~(\ref{general_process}) as
	\begin{equation}
		\frac{d}{dt}\left[ \mathcal{F}(X(t)) \right] = D\mathcal{F}(X)~\partial_t {X} = D\mathcal{F}(X)~P(X)V,
	\end{equation}
	where $D\mathcal{F}(X)$ denotes the derivative of the free energy. We form the Rayleighian  as
	\begin{equation}
		\label{rayleighian}
		\mathcal{R}(V;X) = D\mathcal{F}(X)~P(X)V  -F(X) V + \mathcal{D}(V;X).
	\end{equation}
	Onsager's variational principle then states that the system evolves such that
	\begin{equation}
		\label{Onsagermain}
		V = {\text{argmin}}_{W}~\mathcal{R}(W;X), \;\;\; \mbox{subject to} \;\;\; \mathbb{C}(X) W=0.
	\end{equation}
	The constrained dynamics can be equivalently characterized as stationary points of the Lagrangian
	\begin{equation}
		\mathcal{L}(V,\Lambda;X) = D\mathcal{F}(X)~P(X)V  + \mathcal{D}(V;X)  -F(X) V + \Lambda\cdot\mathbb{C}(X)~V,
	\end{equation}
	where $\Lambda$ are the Lagrange multipliers. Once $V$ is obtained from this variational principle, we can then integrate $\partial_t{X}$ in time recalling Eq.~\eqref{general_process}. 
	
	Let us examine the first-order optimality conditions. The stationarity condition $0 = \delta_\Lambda \mathcal{L}$ simply leads to $0 = \mathbb{C}(X)V$. The stationarity condition $0 = \delta_V \mathcal{L}$ leads to
	\begin{equation}
		\label{weakgen}
		0 = D \mathcal{F}(X)~P(X)  + D_V\mathcal{D}(V;X) -F(X) +  \Lambda\cdot\mathbb{C}(X),
	\end{equation}
	where $D_V\mathcal{D}$ denotes the derivative of dissipation with respect to its first argument. This equation establishes a balance between thermodynamic driving forces, dissipative forces, external/active forces and constraint forces. If $\mathcal{D}$ is smooth, then generalized reciprocal relations are simply the statement of symmetry of second derivatives of $\mathcal{D}$ with respect to different components of $V$ \cite{EDELEN1972481}.
	
	Multiplying Eq.~(\ref{weakgen}) by the actual $V$ along the dynamics, using the fact that $\mathbb{C}(X)V=0$  and rearranging terms, we obtain
	\begin{equation}
		\underbrace{D\mathcal{F}(X)~P(X) V}_{d{\mathcal{F}}/dt} = -D_V\mathcal{D}(V;X) V + F(X)V,
		\label{lya}
	\end{equation}
	which is a statement of energy balance relating the rate of change of the free energy, the power dissipated in irreversible processes, and the external/active power input. For a quadratic dissipation potential, we have $D_V\mathcal{D}(V;X)V = 2 \mathcal{D}(V;X)$ and hence the dissipated power is twice the dissipation potential. 
	
	The second-order optimality condition for $V$ to be a minimum of the Rayleighian is the condition that $\mathcal{D}$ is a convex function of $V$, 
	which leads to
	\begin{equation}
		\mathcal{D}(0;X) \ge \mathcal{D}(V;X) - D_V\mathcal{D}(V;X)V.
	\end{equation}
When supplemented by the natural conditions $\mathcal{D}(0;X) = 0$ and $\mathcal{D}(V;X)\ge 0$, we conclude that  
	\begin{equation}
		\label{entropy_prod}
		D_V\mathcal{D}(V;X)V \ge 0.
	\end{equation}
	This equation is an entropy production inequality for irreversible processes. Hence, the existence of a non-negative and convex dissipation potential satisfying $\mathcal{D}(0;X) = 0$ from which dissipative forces derive is the nonlinear generalization of Onsager's relations and the entropy production inequality \cite{EDELEN1972481,mielke2016generalization}. In the absence of external/active forces, we conclude from Eqs.~(\ref{lya},\ref{entropy_prod}) that $d{\mathcal{F}}/dt = - D_V\mathcal{D}(V;X)V \le 0$, and hence the free energy $\mathcal{F}$ is a Lyapunov function of the dynamics.
	
	In summary, Onsager's variational principle is thermodynamically consistent by construction; the first-order optimality condition establishes energy conservation, and the second-order optimality condition guarantees non-negative entropy production. 	

\section{\mab{Elaboration of the rate-of-change of a frame-indifferent nematic free-energy}}
\label{App:dFdt}

We elaborate next an expression for the time derivative of the free-energy density $f$ appearing in Eq.~(\ref{eq:change_of_free_energy}) in terms of the process variables of the theory. Using the chain rule, the material time derivative of $f$ can be written as
	\begin{eqnarray} 
		\dot{f} =   \frac{\partial f}{\partial q_{ab}} \dot{q}_{ab}+ \frac{\partial f}{\partial \nabla_c q_{ab}} \frac{D}{Dt} \left(\nabla_c q_{ab}\right), \label{eq:dotf}  
	\end{eqnarray}
	where we have introduced the more explicit notation $D/Dt$ for the material time-derivative when required for clarity. To further elaborate on this expression, we note that, unlike partial time and space differentiation, the material time-derivative and $\nabla$ do not commute. Indeed, the material time derivative of $\nabla \bm{q}$ is given by
	\begin{equation}
		\frac{D}{Dt} \left(\nabla_c q_{ab}\right) = \partial_t  \nabla_c q_{ab} + v_d  \nabla_d \nabla_c q_{ab},
	\end{equation}
	whereas the gradient of the material time derivative of $\bm{q}$ is
	\begin{eqnarray}
		\nabla_c \dot{q}_{ab} & = \nabla_c \left(\frac{D}{Dt} {q}_{ab}\right) = \nabla_c \partial_t   q_{ab} + \nabla_c v_d \nabla_d q_{ab} + v_d  \nabla_c \nabla_d  q_{ab} \nonumber \\ & = \frac{D}{Dt} \left(\nabla_c q_{ab}\right) + \nabla_c v_d q_{ab}  = \frac{D}{Dt} \left(\nabla_c q_{ab}\right) + d_{dc} \nabla_d q_{ab} + w_{dc} \nabla_d q_{ab}.\label{aux1}
	\end{eqnarray}
	To express $\dot{f}$ in terms of our process variable $\widehat{\bm{q}}$, we compute its gradient, see Eq.~(\ref{eq:Jaumann}), as
	\begin{eqnarray}
		\nabla_c \widehat{q}_{ab} =  \nabla_c \dot{q}_{ab}  + \nabla_c q_{ad} w_{db} + q_{ad}\nabla_c  w_{db}  + \nabla_c q_{db} w_{da} + q_{db} \nabla_c  w_{da}.
	\end{eqnarray}
	Combining Eq.~(\ref{aux1}) and the definition of $\bm{\zeta}$ in Eq.~(\ref{zeta}), we rewrite this expression as 
	\begin{eqnarray}
		\nabla_c \widehat{q}_{ab} = &  \frac{D}{Dt} \left(\nabla_c q_{ab}\right) + d_{dc} \nabla_d q_{ab} + w_{dc} \nabla_d q_{ab}  +  w_{db} \nabla_c q_{ad} + w_{da} \nabla_c q_{db}  \nonumber \\ & +  \left(q_{ad}\epsilon_{db}   - \epsilon_{ad} q_{db} \right)\zeta_c.
	\end{eqnarray}
	Using this expression, the material time derivative of $f$ in Eq.~(\ref{eq:dotf}) takes the form
	\begin{eqnarray} 
		\dot{f} &= &  \frac{\partial f}{\partial q_{ab}} \left(\widehat{q}_{ab}- q_{ad} w_{db}- q_{db} w_{da}\right)   + \frac{\partial f}{\partial \nabla_c q_{ab}} \bigg[ \nabla_c \widehat{q}_{ab} +  2\epsilon_{ad} q_{db}   \zeta_c \nonumber\\ && \;\;\;\;\; - d_{dc} \nabla_d q_{ab}  - w_{dc} \nabla_d q_{ab}  -  w_{db} \nabla_c q_{ad} - w_{da} \nabla_c q_{db}  \bigg], \label{eq:dotfUUUU}  
	\end{eqnarray}
	where we have used ${q}_{ab} = {q}_{ba}$ and ${\epsilon}_{ab} = -{\epsilon}_{ba}$ to simplify the term involving $\zeta_c$.
	
	The free energy density $f$ should be frame indifferent, and hence its material time derivative should vanish for any rigid body motion characterized by  $d_{ab}=0$, $\widehat{q}_{ab}=0$, and uniform but otherwise arbitrary $w_{ab}$, and hence $\zeta_c = 0$. Invoking this principle along with  Eq.~(\ref{eq:dotfUUUU}), we find that the identity
	\begin{eqnarray} 
		0 &= & \frac{\partial f}{\partial q_{ab}} \left(  q_{ad} w_{db}+ q_{db} w_{da}\right)  + \frac{\partial f}{\partial \nabla_c q_{ab}} \bigg[   w_{dc} \nabla_d q_{ab}  +  w_{db} \nabla_c q_{ad} + w_{da} \nabla_c q_{db}  \bigg],
		\label{f_frame_indiff}
	\end{eqnarray}
	should hold for all antisymmetric tensors $\bm{w}$ and for all fields $\bm{q}$. Combining Eqs.~(\ref{f_frame_indiff}) and (\ref{eq:dotfUUUU}), we finally obtain Eq.~(\ref{eq:final_rate_of_free_energy}).

	Further particularizing the frame indifference condition in Eq.~(\ref{f_frame_indiff}) to uniform nematic fields ($\nabla\bm{q} = \bm{0}$)  and spin tensors of the form $\bm{w} = \bm{\epsilon}$, we obtain the condition 
	\begin{equation} 
		0 =  \frac{\partial f}{\partial q_{ac}} q_{cb} - \frac{\partial f}{\partial q_{bc}} q_{ca}.
		\label{f_frame_indiff_2}
	\end{equation}
	Similarly, considering a point where $\bm{q}=\bm{0}$ but its gradient is not, we obtain
	\begin{equation} 
		0 =  \frac{\partial f}{\partial \nabla_a q_{cd}} \nabla_b q_{cd} -  \frac{\partial f}{\partial \nabla_b q_{cd}} \nabla_a q_{cd} + 	
		2\left(\frac{\partial f}{\partial \nabla_d q_{ac}} \nabla_d q_{cb} - \frac{\partial f}{\partial \nabla_d q_{bc}} \nabla_d  q_{ca}\right).
		\label{f_frame_indiff_3}
	\end{equation}
	Equations (\ref{f_frame_indiff_2},\ref{f_frame_indiff_3}) are thus identities that should hold for all $q_{ab}$ and for all $\nabla_c q_{ab}$, and that express frame indifference of the free energy. These expressions are useful in the next Appendix.

\section{\mab{Elaboration of stress tensor}}
\label{App:stress}

We provide here a more explicit expression for the total Cauchy stress tensor of the general theory. From Eqs.~(\ref{sigma_a}) and (\ref{eq:moment}), we obtain
	\begin{eqnarray}
		{\sigma}_{ab}^{\text{a}}  & = &\nabla_c \left[\rho\left( \frac{\partial f}{\partial \nabla_c q_{bd}}q_{ad} - \frac{\partial f}{\partial \nabla_c q_{ad}}q_{bd}   \right) \right]-\frac{1}{2}\nabla_c \left( \rho \frac{\partial (  d+p)}{\partial \zeta_c}\right)\epsilon_{ab} - \omega_{ab} \nonumber \\
		 & = &  \rho \left( \frac{\partial f}{\partial \nabla_c q_{bd}} \nabla_c q_{ad} -   \frac{\partial f}{\partial \nabla_c q_{ad}} \nabla_c q_{bd}  \right) + q_{ad} \nabla_c\left( \rho  \frac{\partial f}{\partial \nabla_c q_{bd}}\right) -  q_{bd} \nabla_c\left( \rho  \frac{\partial f}{\partial \nabla_c q_{ad}}\right) \nonumber \\
		 && -\frac{1}{2}\nabla_c \left( \rho \frac{\partial (  d+p)}{\partial \zeta_c}\right)\epsilon_{ab} - \omega_{ab} \nonumber \\
		 & = &  - \frac{\rho}{2}  \left( \frac{\partial f}{\partial \nabla_b q_{cd}} \nabla_a q_{cd} -   \frac{\partial f}{\partial \nabla_a q_{cd}} \nabla_b q_{cd}  \right) + q_{ad} h_{bd} -  q_{bd} h_{ad} \nonumber \\
		 && -\frac{1}{2}\nabla_c \left( \rho \frac{\partial (  d+p)}{\partial \zeta_c}\right)\epsilon_{ab} - \omega_{ab} \label{eq::explicit_antisymmetric_stress}
\end{eqnarray}
where in the last step we have invoked frame-indifference of $f$ as expressed by Eqs.~(\ref{f_frame_indiff_2}) and (\ref{f_frame_indiff_3}) and the definition of the generalized nematic force $\bm{h}$ in Eq.~(\ref{nem_field}). Adding Eqs. (\ref{eq::explicit_antisymmetric_stress}) and 	(\ref{eq:sym_stress}), we obtain Eq.~(\ref{total_stress}).

			\section{Conditions for non-negative entropy production} \label{inequality}
			
			We identify here the conditions for non-negative entropy production. It is obvious that $\mathcal{D}\left[\bm{0},\bm{0}\right]=0$. We thus examine when $\mathcal{D}$ is non-negative and convex. The integrand $d$ can be written as 
			\begin{eqnarray}
				d(\bm{v},\bm{d},\widehat{\bm{q}}) &  = \eta \big( d_{ab} d_{ab}   + (\text{tr}d)^2  \big) + \frac{\eta_{\text{rot}}}{2} \widehat{q}_{ab} \widehat{q}_{ab}  + \beta d_{ab}^{\rm dev}\widehat{q}_{ab} + \frac{\gamma}{2} v_a v_a \geq 0  \nonumber\\ 
				& = 2\eta \big( d_{11}^2 + d_{12}^2 + d_{22}^2  + d_{11}d_{22} \big) + \eta_{\text{rot}} \big( \widehat{q}_{1}^{\, \, 2} + \widehat{q}_{2}^{\, \, 2}  \big)  \\ 
				&  + \beta \big(d_{11}\widehat{q}_{1} - d_{22}\widehat{q}_{1} + \nonumber  2d_{12}\widehat{q}_{2}\big) +  \nonumber  \frac{\gamma}{2} \big(v_1^{\,2} + v_2^{\,2}\big) \geq 0 .
			\end{eqnarray}
			and hence it is a quadratic form of its arguments that can be expressed as $d = z_{a} M_{ab} z_b$ with 
			\begin{equation}
				\bm{z} = \left(\begin{array}{c}
					d_{11}\\ 
					d_{22}\\ 
					d_{12}\\ 
					q_{1}\\ 
					q_{2}\\ 
					v_1\\ 
					v_2\\ 
				\end{array}\right), \quad  \bm{M} = \left(\begin{array}{ccccccc}
					2\eta& \eta &0  &\beta/2  &0  &0  &0   \\ 
					\eta& 2\eta &0  &-\beta/2  &0  &0  &0   \\ 
					0&  0&  2\eta&  0& \beta &0  &0   \\ 
					\beta/2& -\beta/2 &0  &\eta_{\rm rot}  &0  &0  &0   \\ 
					0&  0 &\beta  & 0 & \eta_{\rm rot} &0  &0   \\ 
					0& 0 & 0 & 0 & 0 & \gamma/2 & 0  \\ 
					0& 0 & 0 & 0 & 0 & 0 & \gamma/2
				\end{array}\right).
			\end{equation}
			Because $\bm{z}$ is a linear function of $\left(\bm{v},\widehat{\bm{q}}\right)$, a direct argument shows that if the symmetric matrix $\bm{M}$ is positive semi-definite, then $\mathcal{D}\left[\bm{v},\widehat{\bm{q}}\right]$ is a non-negative and convex functional. According to Sylvester's criterion \cite{doi:10.1080/00029890.1991.11995702}, $\bm{M}$ is positive semi-definite if and only if its leading principal minors are non-negative. As simple calculation shows that, since $\eta>0$, $\eta_{\rm rot}>0$ and $\gamma>0$, this condition is met when
			\begin{equation}
				2\eta \eta_{\rm rot} - \beta^2 \geq 0.
			\end{equation}

	\section{Finite element formulation} \label{sec_4}

	The governing Eqs.~(\ref{first_govern}-\ref{last_govern}) of the proposed model are non-linear and involve tight couplings between nematic, velocity and density fields. For this reason, a solution of the governing equations in arbitrary geometries and boundary conditions cannot be obtained by analytical means. Here, we develop a finite element computational approach building on Onsager's variational formalism. In this setting, numerical space discretization is straightforward and follows from performing extremization of the Lagrangian in a constrained functional space given by the finite element approximation of the process variable fields. The resulting stationarity conditions represent the discretized weak form of the governing equations. For time discretization, we resort to the implicit Euler method. 
	
	\subsection{Weak form of the governing equations}
	
	In Eqs.~(\ref{eq:Rayleighian}) and~(\ref{eq:Lagrangian}), we have expressed the Rayleghian and the Lagrangian in terms of the independent variables $\dot{\rho},\widehat{\bm{q}},\bm{v},\bm{d},\bm{w}$ and $\bm{\zeta}$ to derive the governing equations in the most physically meaningful form. However, to derive an Eulerian finite element method, it is more convenient to  consider $\partial_t{\bm{q}}$ and $\bm{v}$ as the sole process variables. From the first expression in Eq.~(\ref{eq:change_of_free_energy}), applying the chain rule to compute $\partial_t f$ and using Eq.~(\ref{eq:balance_mass}), we can express the rate of change of the free energy as
	\begin{eqnarray} 
		\frac{d{\mathcal{F}}}{dt}\left[\partial_t\bm{q},\bm{v};\rho,\bm{q}\right] = &  \int_A \left\{ \rho \frac{\partial f}{\partial \bm{q}} : \partial_t \bm{q} + \rho \frac{\partial f}{\partial \nabla_c {q}_{ab}}  \nabla_c \partial_t {q}_{ab} + f \left[r-\nabla\cdot\left(\rho\bm{v}\right)\right] \right\}dA  \nonumber\\ 
		&  +  \int_{\partial_{N}A} f\rho \bm{v} \cdot\bm{N} dl. \label{eq:discrete_free_energy}
	\end{eqnarray}
	For the dissipation and power potentials, we directly substitute the definitions of $\bm{w}$ and $\bm{\zeta}$ as a function of gradients of $\bm{v}$ and use Eq.~(\ref{eq:Jaumann}) to write $\widehat{\bm{q}}$ in terms of $\partial_t \bm{q}$ and $\bm{v}$, formally leading to functionals of the form $\mathcal{D}[\partial_t\bm{q},\bm{v};\rho,\bm{q}]$ and $\mathcal{P}[\partial_t\bm{q},\bm{v};\rho,\bm{q}]$.
	%\begin{eqnarray}   \label{eq:discrete_dissipation}
	%	\mathcal{D}[\partial_t\bm{q},\bm{v};\rho,\bm{q}] = \int_A d\left(\partial_t\bm{q},\bm{v};\bm{q}\right)  \rho dA,
	%\end{eqnarray}
	%\begin{eqnarray}  \label{eq:discrete_power}
	%	\mathcal{P}[\partial_t\bm{q},\bm{v};\rho,\bm{q}] = \int_A p\left(\partial_t\bm{q},\bm{v};\bm{q}\right)  \rho dA.
	%\end{eqnarray}
	Combining these functionals, the Rayleighian can be expressed as
	\begin{equation}  \label{eq:ray_discrte}
		\mathcal{R}[\partial_t\bm{q},\bm{v};\rho,\bm{q}] = \frac{d{\mathcal{F}}}{dt}[\partial_t\bm{q},\bm{v};\rho,\bm{q}]+ \mathcal{D}[\partial_t\bm{q},\bm{v};\rho,\bm{q}]+ \mathcal{P}[\partial_t\bm{q},\bm{v};\rho,\bm{q}].
	\end{equation}
	Onsager's variational principle then provides an alternative form of the governing equations by minimizing this Rayleighian with respect to $\partial_t \bm{q}$ and $\bm{v}$.  Minimization with respect to  $\partial_t\bm{q}$ leads to 
	\begin{eqnarray}
		0&=\delta_{\partial_t \bm{q}} \mathcal{R} 
		&=  \int_A  \bigg[ \left(\frac{\partial  f}{\partial \bm{q}} +   \frac{\partial (d+p)}{\partial \widehat{\bm{q}}}  \right):\bm{p} + \frac{\partial f}{\partial \nabla_c {q}_{ab}}  \nabla_c  {p}_{ab} \bigg]\rho dA -   \int_{\partial_{N_{\bm{L}}}A} \bm{L}:\bm{p} dl, \label{eq:weak_q_disc}
	\end{eqnarray}
	where $\bm{p}$ is an arbitrary variation of $\partial_t \bm{q}$, and hence a traceless and symmetric second order tensor. Integration by parts of this weak form to obtain the corresponding Euler-Lagrange equations is not required for the finite element discretization. 
	
	Minimization with respect to $\bm{v}$ leads to
	\begin{eqnarray}
		\bm{0}=\delta_{\bm{v}} \mathcal{R}=   ~\int_A &\left\{-\frac{ f}{\rho} \nabla \cdot \left(\rho\bm{u}\right) + \frac{\partial ( d+ p)}{\partial \bm{v}} \cdot\bm{u}    \right.  
		+ \left[\frac{\partial ( d+ p)}{\partial \bm{d}} + \frac{\partial ( d+ p)}{\partial \bm{w}}\right] : \nabla\bm{u}\nonumber\\
		&+ \left. \frac{\epsilon_{ab}}{2} \frac{\partial ( d+ p)}{\partial {\zeta}_c} \nabla_c \nabla_b {u}_a + \frac{\partial ( d+ p)}{\partial \widehat{\bm{q}}} : \delta_{\bm{v}} \widehat{\bm{q}} \right\} \rho dA 
		+   \int_{\partial A} f\rho \bm{N}\cdot\bm{u}\,dl	
		\nonumber\\
		& - \int_{\partial_{N_{\bm{t}}} A} \bm{t} \cdot \bm{u} \,dl -  \int_{\partial_{N_{\bm{\Gamma}}} A} \bm{\Gamma}:\nabla\bm{u} \,dl -  \int_{\partial_{N_{\bm{L}}} A} \bm{L}:\delta_{\bm{v}} \widehat{\bm{q}} \,dl
		\label{eq:weak_v_disc}
	\end{eqnarray}  
	where $\bm{u}$ is an arbitrary variation of $\bm{v}$. The variation of the Jaumann derivative of the nematic order tensor with respect to velocity is given by $\delta_{\bm{v}} \widehat{{q}}_{ab}=  {u}_c\nabla_c{q}_{ab} + \frac{1}{2}\left({q}_{ac} {\epsilon}_{cb} -{\epsilon}_{ac}{q}_{cb} \right) {\epsilon}_{ef}\nabla_f{u}_e$. 
	
We note that, although not written here, non-homogeneous Dirichlet boundary conditions on $\partial_{D_{\bm{w}}} A $ and $\partial_{D_{\widehat{\bm{q}}}} A $ would need to be explicitly enforced in the formulation presented here, e.g.~through Lagrange multipliers, because constraints on $\widehat{\bm{q}}$ and $\bm{w}$ depend on  $\partial_t \bm{q}$ and $\bm{v}$ in a non-trivial manner. 
	
The equivalence between Eq.~(\ref{eq:weak_v_disc}) and the form of balance of linear momentum found earlier in Eq.~(\ref{eq:weak_v}) is not obvious even if both of these equations encode the same physics. In \ref{equivalence}, we explicitly show this equivalence. 

For balance of mass, we consider the weak form of Eq.~(\ref{eq:balance_mass}) by  multiplying by an arbitrary test function $\delta\rho$, integrating, and applying the divergence theorem to the diffusive term to obtain
	\begin{eqnarray}
		\int_A \left \{\left(  \frac{\partial \rho}{\partial t} +  \nabla \cdot (\rho \bm{v}) - k_p  +\rho k_d\right) \delta \rho  + D\nabla \rho \cdot \nabla \delta \rho \right \} dA \nonumber \\ 	-  \int_{\partial A} D  \nabla \rho \cdot \bm{N} \delta \rho  dl  = 0.  \label{eq:weak_mass}
	\end{eqnarray}
	The boundary integral is dealt with by either prescribing the diffusive flux across the boundary or by considering $\delta\rho=0$ in parts of the boundary where $\rho$ is fixed.
	
	\subsection{Space discretization}

	We discretize fields in space following a typical finite element approximation based on a mesh with $N$ vertices. Each vertex or node $I$ of the mesh has an associated basis function $B_I(\bm{x})$. The choice of the type of basis function depends on the required regularity. The weak form in Eq.~(\ref{eq:weak_v_disc}) contains second-order derivatives of $\bm{u}$, and hence should require basis functions with at least square-integrable second order derivatives. However, since in our model $\partial ( d+ p) / \partial \bm{\zeta} = \bm{0}$, only first order derivatives of $\bm{u}$ (and $\bm{v}$) appear, and hence a conventional finite element discretization with $C^0$ continuity can be used. The density field is discretized as
	\begin{equation}
		\rho(\bm{x},t) = \sum_{I=1}^N \rho_I(t) B_I(\bm{x}), \label{eq:shape_func}
	\end{equation}
	where $\rho_I(t)$ is the $I-$th nodal coefficient at time $t$. Analogously, we have 
	\begin{equation}
		\bm{v}(\bm{x},t) = \sum_{I=1}^N\bm{v}_I(t) B_I(\bm{x}),
	\end{equation}
	where the nodal degrees of freedom are vectors. 
	We represent the traceless and symmetric nematic tensor field as
	\begin{equation}
		\bm{q}(\bm{x},t) = \left(\begin{array}{cc}
			q_1(\bm{x},t) & q_2(\bm{x},t)\\
			q_2(\bm{x},t) & -q_1(\bm{x},t)
		\end{array}\right),
	\end{equation}
	and discretize its components as
	\begin{equation}
		q_1(\bm{x},t) = \sum_{I=1}^N q_{1I} (t) B_I(\bm{x}), \qquad  q_2(\bm{x},t) = \sum_{I=1}^N q_{2I} (t) B_I(\bm{x}).
	\end{equation}
	Thus, we have 5 degrees of freedom per node, namely $\rho_I, v_{1I}, v_{2I}, q_{1I}$ and $q_{2I}$.
	
	To discretize Eq.~(\ref{eq:weak_q_disc}), we express variations $\bm{p}$ as linear combinations of the following traceless symmetric tensors 
	\begin{equation}
		\left(\begin{array}{cc}
			B_I(\bm{x}) & 0\\
			0 & -B_I(\bm{x})
		\end{array}\right) \qquad \mbox{and} \qquad  \left(\begin{array}{cc}
			0 & B_I(\bm{x})\\
			B_I(\bm{x}) & 0
		\end{array}\right),
	\end{equation}
	for all $I$.  Using the space-discretized nematic tensor and the variations defined above, the discretized form of balance of generalized force conjugate to nematic order in Eq.~(\ref{eq:weak_q_disc}) becomes a set of $2N$  algebraic equations. For the balance of linear momentum, Eq.~(\ref{eq:weak_v_disc}), we consider variations of velocity $\bm{u}$ to be linear combinations of $[B_I(\bm{x})\;\; 0]^T$ and $[0\;\; B_I(\bm{x})]^T$ to obtain $2N$ additional algebraic equations. Finally, for the balance of mass, Eq.~(\ref{eq:weak_mass}), we consider $\delta\rho$ to be linear combinations of $B_I(\bm{x})$ to obtain $N$ equations. We note that when advection dominates in Eq.~(\ref{eq:weak_mass}), such a Galerkin approach, in which $\delta\rho$ are discretized with the same basis functions as $\rho$, leads to numerical instabilities. In our implementation, we check the condition for stability and stabilize the numerical formulation using the SUPG method if required \cite{donea2003}.  Hence, we obtain a set of $5N$ differential-algebraic equations involving $\rho_I$, $v_{1I}$, $v_{2I}$, $q_{1I}$, $q_{2I}$, and the time-derivatives of the density and nematic degrees of freedom.

	\subsection{Time discretization and solution method}

	%Together,  Eqs.~(\ref{eq:weak_mass},\ref{eq_week_1}-\ref{eq_week_4}) constitute a differential-algebraic system of $5N$ coupled equations. 
	We discretize these equations in time using a non-uniform grid of time-steps that we denote by the superindex $[\text{n}]$. We denote by $\rho^{[\text{n}]}_I$, $\bm{q}^{[\text{n}]}_I$, and $\bm{v}^{[\text{n}]}_I$ the nodal coeficients at the $n-$th time-step, and consider  a backward Euler approximation, according to which we evaluate all fields in Eqs.~(\ref{eq:weak_q_disc}-\ref{eq:weak_mass}) at step $[\text{n}]$ and approximate time-derivatives as $\partial_t \rho_I^{[\text{n}]} = (\rho^{[\text{n}]}_I-\rho^{[\text{n-1}]}_I)/\Delta t^{[\text{n}]}$, $\partial_t \bm{q}_I^{[\text{n}]} = (\bm{q}^{[\text{n}]}_I-\bm{q}^{[\text{n-1}]}_I)/\Delta t^{[\text{n}]}$. This leads to a set of nonlinear algebraic equations that we solve with Newton-Raphson's method. 
	
			\subsection{Numerical convergence} \label{appendix_grid}

\begin{figure}[t]
				\centering
				\includegraphics[width=0.9\textwidth]{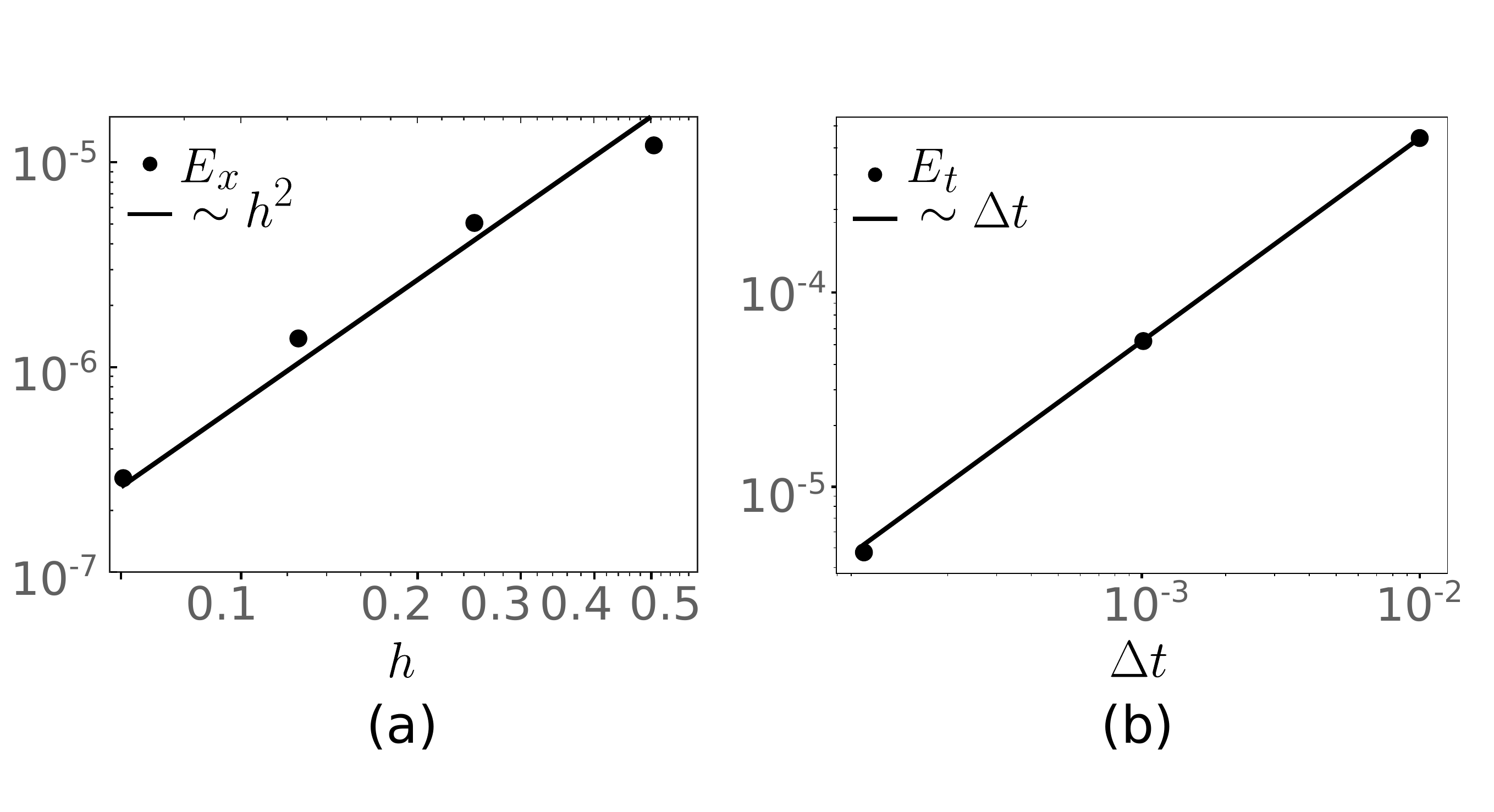}
				\caption{\label{grid} Convergence of the numerical method. (a) Convergence of the spatial discretization. The relative free-energy error $E_x$  shows a convergence rate $\sim h^{2}$. (b) Convergence of the temporal discretization. The relative free-energy error  $E_t$  shows a convergence rate $\sim \Delta t$.}
			\end{figure}

			We verify our numerical methods for the spatial discretization by solving the proposed model for various mesh sizes. For this purpose, we consider a passive nematic system with circular confinement, see Fig.~\ref{sec_1_chap_3_fig_3}, and $\ell_p/\ell_0=0.04$. We define the normalized error of free-energy at steady state
			\begin{equation}	
				E_x = \frac{|F^h -F^*|}{|F^*|} , 
			\end{equation}
			where $F^h$ is the free-energy of a finite element solution with mesh-size $h$ and $F^*$ is that of an overkill solution computed with a very fine mesh of $n_e = 695,296$ triangular elements. We plot the energy error for decreasing mesh sizes, Fig.~\ref{grid}(a). The results show the expected optimal convergence (slope of 2 in a log-log scale) for linear elements. 
			
			Next for the same numerical experiment, we validate the temporal convergence by gradually decreasing the
			time step $\Delta t$ used to reach a fixed time point $t|a|/\eta_{\rm rot}=1$. We calculate the relative error of the time-discretization as
			\begin{equation}	
				E_t = \frac{|F^{\Delta t} -F^{t,*}|}{|F^{t,*}|} , 
			\end{equation}
			where $F^{\Delta t}$ is the free-energy obtained with time-step $\Delta t$  at $t|a|/\eta_{\rm rot}=1$ and   $F^{t,*}$ is the free-energy with a very small time-step $\Delta t^* = 10^{-5}$ . As expected,  Fig.~\ref{grid}(b), $E_t$ converges as $\sim \Delta t$.
			
			\section{Equivalence between different forms of balance of linear momentum} \label{equivalence}
			
In this section,  establish an equivalence between the governing equations for balance of linear momentum obtained in Sections \ref{sec_2} and \ref{sec_4}. Comparing Eqs.~(\ref{eq:weak_v}) and (\ref{eq:weak_v_disc}), we need to prove that 
						\begin{eqnarray}
				C_1 = &\int_A \left\{-\frac{ f}{\rho} \nabla_d \left(\rho u_d\right) + \frac{\epsilon_{ab}}{2} \frac{\partial ( d+ p)}{\partial {\zeta}_c} \nabla_c \nabla_b {u}_a  \right. \nonumber \\ & \left. + \frac{\partial ( d+ p)}{\partial \widehat{q}_{ab}}  \left[ {u}_c\nabla_c{q}_{ab} + \frac{1}{2}\left({q}_{ac} {\epsilon}_{cb} -{\epsilon}_{ac}{q}_{cb} \right) {\epsilon}_{ef}\nabla_f{u}_e\right] \right\}\rho dA \nonumber\\
				&+  \int_{\partial A} f\rho N_c u_c dl - \int_{\partial_{N_{\bm{L}}} A} L_{ab} \left[ {u}_c\nabla_c{q}_{ab} + \frac{1}{2}\left({q}_{ac} {\epsilon}_{cb} -{\epsilon}_{ac}{q}_{cb} \right) {\epsilon}_{ef}\nabla_f{u}_e\right]dl \nonumber\\& -\int_{\partial_{N_{\bm{\Gamma}}} A} \Gamma_{ab} \nabla_b u_a dl , \label{eq:weak_v_disc_app}
\end{eqnarray}
 is equal to 
\begin{equation}
				C_2 = \int_A \bar{\sigma}_{ab}\nabla_b u_a dA, \label{eq:weak_v_app}
			\end{equation}
where $\bar{\sigma}_{ab}$ is the total stress except for the terms involving $\partial(d+p)/\partial\bm{d}$ and $\partial(d+p)/\partial\bm{w}$, and it is given by
\begin{eqnarray}
				\bar{\sigma}_{ab} = & -\rho\frac{\partial  f}{\partial \nabla_b q_{dc}} \nabla_a q_{dc}  +   q_{ad}  \nabla_c \left(\rho \frac{\partial f}{\partial \nabla_c {q}_{bd}}\right) - q_{bd}\nabla_c \left(\rho \frac{\partial f}{\partial \nabla_c {q}_{ad}}\right) \nonumber \\ &  -  \frac{1}{2} \nabla_c \left(\rho\frac{\partial (  d+p)}{\partial \zeta_c} \right) \epsilon_{ab}   .
	\end{eqnarray}

Integrating by parts the first term, and accounting for balance of generalized force in Eq.~\eqref{eq:balance_q} to substitute $\partial(d+p)/\partial\widehat{\bm{q}}$, we have 
\begin{eqnarray}
C_1 = &\int_A \left\{ u_d \nabla_d f + \frac{\epsilon_{ab}}{2} \frac{\partial ( d+ p)}{\partial {\zeta}_c} \nabla_c \nabla_b {u}_a  \right. \nonumber \\ & \left. + \left[\frac{1}{\rho}\nabla_d \left(\rho \frac{\partial f}{\partial \nabla_d q_{ab}} \right) - \frac{\partial f}{\partial q_{ab}}\right] \left[ {u}_c\nabla_c{q}_{ab} + \frac{1}{2}\left({q}_{ac} {\epsilon}_{cb} -{\epsilon}_{ac}{q}_{cb} \right) {\epsilon}_{ef}\nabla_f{u}_e\right] \right\}\rho dA \nonumber\\
&- \int_{\partial_{N_{\bm{L}}} A} L_{ab} \left[ {u}_c\nabla_c{q}_{ab} + \frac{1}{2}\left({q}_{ac} {\epsilon}_{cb} -{\epsilon}_{ac}{q}_{cb} \right) {\epsilon}_{ef}\nabla_f{u}_e\right]dl \nonumber\\& -\int_{\partial_{N_{\bm{\Gamma}}} A} \Gamma_{ab} \nabla_b u_a dl.
\end{eqnarray}
Integrating by parts the first term within square brackets in the second line of this equation, we obtain
\begin{eqnarray}
C_1 = &\int_A \left\{ u_d \nabla_d f + \frac{\epsilon_{ab}}{2} \frac{\partial ( d+ p)}{\partial {\zeta}_c} \nabla_c \nabla_b {u}_a  \right. \nonumber \\  
& -  \frac{\partial f}{\partial \nabla_d q_{ab}} \nabla_d\left[ {u}_c\nabla_c{q}_{ab} + \frac{1}{2}\left({q}_{ac} {\epsilon}_{cb} -{\epsilon}_{ac}{q}_{cb} \right) {\epsilon}_{ef}\nabla_f{u}_e\right]  \nonumber\\
&\left.  - \frac{\partial f}{\partial q_{ab}} \left[ {u}_c\nabla_c{q}_{ab} + \frac{1}{2}\left({q}_{ac} {\epsilon}_{cb} -{\epsilon}_{ac}{q}_{cb} \right) {\epsilon}_{ef}\nabla_f{u}_e\right] \right\}\rho dA \nonumber\\
&- \int_{\partial_{N_{\bm{L}}} A}\left( L_{ab} - \rho \frac{\partial f}{\partial \nabla_d q_{ab}} N_d \right)\left[ {u}_c\nabla_c{q}_{ab} + \frac{1}{2}\left({q}_{ac} {\epsilon}_{cb} -{\epsilon}_{ac}{q}_{cb} \right) {\epsilon}_{ef}\nabla_f{u}_e\right]dl \nonumber\\& -\int_{\partial_{N_{\bm{\Gamma}}} A} \Gamma_{ab} \nabla_b u_a dl.
\end{eqnarray}
The boundary integral over $\partial_{N_{\bm{L}}} A$ vanishes because of the boundary condition in Eq.~\eqref{eq:balance_q_bd}. Noting that 
\begin{equation}
\nabla_d f = \frac{\partial f}{\partial q_{ab}}\nabla_d q_{ab} + \frac{\partial f}{\partial \nabla_c q_{ab}} \nabla_c \nabla_d q_{ab}, 
\end{equation}
we can cancel three terms in the first three lines of this equation to obtain
\begin{eqnarray}
C_1 = &\int_A \left\{ \frac{\epsilon_{ab}}{2} \frac{\partial ( d+ p)}{\partial {\zeta}_c} \nabla_c \nabla_b {u}_a  \right. \nonumber \\  
& -  \frac{\partial f}{\partial \nabla_d q_{ab}} \nabla_d\left[  \frac{1}{2}\left({q}_{ac} {\epsilon}_{cb} -{\epsilon}_{ac}{q}_{cb} \right) {\epsilon}_{ef}\nabla_f{u}_e\right]  -  \frac{\partial f}{\partial \nabla_d q_{ab}} \nabla_d {u}_c\nabla_c{q}_{ab} \nonumber\\
&\left.  - \frac{\partial f}{\partial q_{ab}} \left[  \frac{1}{2}\left({q}_{ac} {\epsilon}_{cb} -{\epsilon}_{ac}{q}_{cb} \right) {\epsilon}_{ef}\nabla_f{u}_e\right] \right\}\rho dA \nonumber\\
\nonumber\\& -\int_{\partial_{N_{\bm{\Gamma}}} A} \Gamma_{ab} \nabla_b u_a dl.
\end{eqnarray}
Because of frame indifference of $f$ as expressed by Eq.~(\ref{f_frame_indiff_2}), the term in the third line vanishes. Furthermore, using the symmetry of $\bm{q}$, we can simplify the second line as
\begin{eqnarray}
C_1 = &\int_A \left\{ \frac{\epsilon_{ab}}{2} \frac{\partial ( d+ p)}{\partial {\zeta}_c} \nabla_c \nabla_b {u}_a  \right. \nonumber \\  
&\left.  -  \frac{\partial f}{\partial \nabla_d q_{ab}} \nabla_d \left(  {q}_{ae} \nabla_b {u}_e -  {q}_{ae} \nabla_e {u}_b\right)  -  \frac{\partial f}{\partial \nabla_d q_{ab}} \nabla_d {u}_c\nabla_c{q}_{ab} \right\}\rho dA \nonumber\\
\nonumber\\& -\int_{\partial_{N_{\bm{\Gamma}}} A} \Gamma_{ab} \nabla_b u_a dl.
\end{eqnarray}
 Then, integration by parts of the term in the first line and the first term in the second line yields
\begin{eqnarray}
C_1 = &\int_A \left\{-\frac{1}{2}\nabla_c\left(\rho  \frac{\partial ( d+ p)}{\partial {\zeta}_c}\right) \epsilon_{ab} \nabla_b {u}_a  \right. \nonumber \\  
& \left. +   \nabla_d\left( \rho\frac{\partial f}{\partial \nabla_d q_{ab}}\right) \left(  {q}_{ae} \nabla_b {u}_e -  {q}_{ae} \nabla_e {u}_b\right)  -  \rho\frac{\partial f}{\partial \nabla_d q_{ab}} \nabla_d {u}_c\nabla_c{q}_{ab} \right\} dA \nonumber\\
& -\int_{\partial_{N_{\bm{\Gamma}}} A} \left( \Gamma_{ab} -  \frac{1}{2} \rho  \frac{\partial ( d+ p)}{\partial {\zeta}_c} N_c \epsilon_{ab} \right)\nabla_b u_a dl \nonumber \\
& -\int_{\partial_{N_{\bm{\Gamma}}} A}   \rho\frac{\partial f}{\partial \nabla_d q_{ab}} N_d \left(  {q}_{ae} \nabla_b {u}_e -  {q}_{ae} \nabla_e {u}_b\right)  dl.
\end{eqnarray}
Renaming the dummy indices conveniently, we obtain
\begin{eqnarray}
C_1 = &\int_A \left\{-\frac{1}{2}\nabla_c\left(\rho  \frac{\partial ( d+ p)}{\partial {\zeta}_c}\right) \epsilon_{ab}  \right. \nonumber \\  
& \left. +  {q}_{ae} \nabla_d\left( \rho\frac{\partial f}{\partial \nabla_d q_{eb}}\right)   -    {q}_{be}\nabla_d\left( \rho\frac{\partial f}{\partial \nabla_d q_{ea}}\right)    -  \rho\frac{\partial f}{\partial \nabla_b q_{ef}} \nabla_a{q}_{ef} \right\} \nabla_b {u}_a  dA \nonumber\\
& -\int_{\partial_{N_{\bm{\Gamma}}} A} \left\{ \Gamma_{ab} -  \left[\frac{1}{2} \rho  \frac{\partial ( d+ p)}{\partial {\zeta}_c} \epsilon_{ab} - {q}_{ae} \left( \rho\frac{\partial f}{\partial \nabla_c q_{eb}}\right)   +   {q}_{be}\left( \rho\frac{\partial f}{\partial \nabla_c q_{ea}}\right) \right] N_c \right\}\nabla_b u_a dl.\nonumber
\end{eqnarray}
Recalling the boundary condition in Eq.~(\ref{eq:balance_angular_momentum}), the boundary term vanishes. Finally, a direct comparison of the bulk term with Eq.~(\ref{eq:weak_v_app}) shows that $C_1 = C_2$.

			\section{Parameters used for the numerical studies}

			\begin{table}[H]
				\scriptsize
				\begin{tabular}{l|cc} \textbf{Parameter} &  Fig.~\ref{sec_1_chap_3_fig_1}    &  Fig.~\ref{sec_1_chap_3_fig_2}                \\ \hline 
					$\rho_0$ [\si{\micro \meter}]  & 0.2 & 0.2  \\ 
					$\eta$ [\si{\pascal \second}] & $10^4$ & $10^4$  \\ 
					$\ell_s = \sqrt{\eta/\gamma}$ [\si{\micro \meter]} &10 &10 \\
					$k_d$ [\si{\second}$^{-1}$] &0.1 & 0.1 \\
					$D$ [\si{\micro \meter}$^2$ s$^{-1}$] & 0.02& 0.02  \\
					$\eta_{\rm rot} / \eta$ & 1&1 \\
					$\beta / \eta$ & -1 & -1,-0.1  \\
					$a/\eta$ $[\si{\second}^{-1}]$ & 1 &1   \\
					$b/\eta$ $[\si{\second}^{-1}]$  &4 &4   \\
					$L/\eta$ $[\si{\micro \meter}^2  \si{\second}^{-1}$] & $ 0.01$&$ 0.01$   \\
					$\rho_0 \lambda_{\bigodot}^0 /(2a)$ &0.4  &0.2,0.4,0.6    \\
					$\lambda^0/\eta \, [\si{\second}^{-1}]$ &0.12& 0.12  \\
					$\delta \lambda/ \lambda$   & 5& 5\\
					$\kappa = \lambda_{\rm aniso}/\lambda$  & [0.5,2]  &1.0      \\
					$\ell_0 $  [\si{\micro \meter}]  & 20  & 20  \\
					$\ell_p = \sqrt{L/(2|a|-\rho_0\lambda_{\odot})}$   [\si{\micro \meter}] & 0.056 &  0.0527,0.056,0.06   \\
					\hline
				\end{tabular}
				\caption{Parameters used in the cell wound healing study.}
				\label{modal_parameters_cell_healing}
			\end{table}
			
			\begin{table}[H]
				\scriptsize
				\begin{tabular}{l|cc} \textbf{Parameter}       &  Fig.~\ref{sec_1_chap_3_fig_3}      &  Fig.~\ref{sec_1_chap_3_fig_3.5}         \\ \hline 
					$\eta$ [\si{\pascal \second}]  & $10^4$ & $10^4$  \\ 
					$\ell_s/\ell_0 = \sqrt{\eta/(\gamma \ell_0^2)}$  &10 &10 \\
					$\eta_{\rm rot} / \eta$ & 1&1\\
					$\beta / \eta$ & -0.2 & -0.2 \\
					$a/\eta$ $[\si{\second}^{-1}]$& -1 &-1 \\
					$b/\eta$ $[\si{\second}^{-1}]$   & 2 &2  \\
					$L/(\eta \ell_0^2)$ $[ \si{\second}^{-1}$] & [$8 \times 10^{-4}$ ,$0.18  $ ] & $8 \times 10^{-4}$  \\
					$\lambda_{\rm aniso}/\eta [s^{-1}]$    & 0 &$0$,$8.9 \times 10^{-3}$,$0.08$,$0.32$     \\
					$\ell_a / \ell_0 = \sqrt{L/\left(\lambda_{\rm aniso}S_0\right)}$   & $\infty$  & $\infty, 0.3,0.1,0.05$ \\
					$\ell_p / \ell_0 = \sqrt{L/2|a|}$   &  [0.02, 0.3]  & 0.02 \\
					\hline
				\end{tabular}
				\caption{Parameters used in the confined cell colony study.}
				\label{modal_parameters_cell_colony}
			\end{table}

			\section{Movie captions}
			\qquad 	\textbf{Movie 1} \quad Dynamics during wound healing  for various degrees of anisotropic activity quantified by $\kappa$. The density and nematic fields are represented by colormaps. The direction and length of black arrows indicate direction and magnitude of the velocity field $\bm{v}$. The direction and length of red segments indicate the average molecular orientation $\bm{n}$ and the nematic order $S$.
			
			\textbf{Movie 2} \quad Dynamics of the compressible model in a confined active nematic system for low and intermediate activities (quantified by the ratio between the active nematic length-scale $\ell_a$ and system size $\ell_0$).  The velocity modulus and nematic fields are represented by colormaps. The direction and length of black arrows indicate direction and magnitude of the velocity field $\bm{v}$. The direction and length of red segments indicate the average molecular orientation $\bm{n}$ and the nematic order $S$.
			
			\textbf{Movie 3} \quad	Dynamics of the compressible model in a confined active nematic system for high activities.

				\textbf{Movie 4} \quad	Dynamics of the incompressible model in a confined active nematic system for low, intermediate and high activities.

			\section*{Bibliography}

			\providecommand{\newblock}{}

		\end{document}